\documentclass[journal,draftcls,onecolumn,12pt,twoside]{IEEEtran}

\usepackage{booktabs}
\usepackage{amsmath}
\usepackage{amsthm}
\usepackage{mathrsfs}
\usepackage{mathtools}
\usepackage{graphicx}
\usepackage{subfig}
\usepackage{latexsym}
\usepackage{amssymb,amsbsy,verbatim,array}
\usepackage{epsfig}
\usepackage{epstopdf}
\usepackage{cite}
\usepackage{color}
\usepackage{algpseudocode}
\usepackage{algorithmicx, algorithm}
\usepackage{makecell}
\usepackage{extarrows}
\usepackage{bm}
\usepackage{mathrsfs}
\usepackage{longtable}
\usepackage{float}
\usepackage[normalem]{ulem}
\usepackage{hyperref}

\begin{document}

\title{SCA-LLM: Spectral-Attentive LLM-Based Wireless World Modeling for Agentic Communications}

\author{
Ke He, Le He, Lisheng Fan, Xianfu Lei, Thang X. Vu,~\IEEEmembership{Senior Member,~IEEE}, \\George K. Karagiannidis,~\IEEEmembership{Fellow,~IEEE} and Symeon Chatzinotas,~\IEEEmembership{Fellow,~IEEE}

\thanks{
K. He, T. X. Vu and S. Chatzinotas are with the
Interdisciplinary Centre for Security, Reliability and Trust (SnT), University
of Luxembourg, L-1855 Luxembourg, Luxembourg (e-mail: ke.he@ieee.org, thang.vu@uni.lu, symeon.chatzinotas@uni.lu).

L. He, and L. Fan are both with School of Computer Science of Guangzhou University, Guangzhou 510006, China (e-mail: hele@e.gzhu.edu.cn, lsfan@gzhu.edu.cn).

X. Lei is with the School of Information Science and Technology, Institute of Mobile Communications, Southwest Jiaotong University, Chengdu 610031, China (e-mail: xflei@home.swjtu.edu.cn).

G. K. Karagiannidis is with the Department of Electrical and Computer Engineering, Aristotle University of Thessaloniki, 541 24 Thessaloniki, Greece and also with Cyber Security Systems and Applied AI Research Center, Lebanese American University (LAU), Lebanon (e-mail: geokarag@auth.gr).

% This work was supported in part by the NSFC (Nos. 62271158/62101145),
% and in part by the Natural Science Foundation of Guangdong Province (No.
% 2021A1515011392).

The corresponding author of this paper is Lisheng Fan.

}

}

\maketitle

\begin{abstract}
Future AI-native wireless networks are moving from reactive optimization to agentic decision-making that can sense, predict, and plan under fast-varying channels. This calls for wireless world models that can predict and roll out channel dynamics, for which multi-step channel state information (CSI) prediction offers a practical short-horizon look-ahead. Recent advances in foundation sequence models further motivate large language models (LLMs) as general-purpose dynamics learners when suitably adapted to non-text time-series signals. However, bridging CSI to LLMs is non-trivial because an effective adapter must expose informative spectral and temporal evolution patterns, while prior designs provide limited inductive bias to capture such channel structures. To this end, we propose SCA-LLM, a spectral-attentive LLM-based wireless world modeling framework that bridges CSI to LLMs via a spectral-channel attention (SCA) adapter. Specifically, the SCA adapter performs multi-spectral representation learning to extract informative channel features and align CSI with the LLM’s sequence modeling capability, enabling parameter-efficient adaptation while keeping the LLM backbone largely frozen. Extensive simulations show that SCA-LLM achieves state-of-the-art prediction performance and strong zero-shot generalization, yielding up to $-2.4~\text{dB}$ normalized mean squared error (NMSE) advantage over the previous LLM based method. Our ablation studies further confirm the effectiveness of the proposed SCA adapter in mitigating domain mismatch.

% The recent success of large language models (LLMs) has inspired growing interest in exploring potential applications in wireless communications, especially for channel prediction tasks. However,  domain mismatch issue can arise from their text-based pre-training nature when applying LLMs to channel prediction directly. To mitigate this, the ``adapter + LLM" paradigm has emerged, where an adapter is designed to bridge the domain gap between the channel state information (CSI) data and LLMs. While initial success demonstrated, existing adapters not fully exploit the potential of this paradigm. Therefore, this work aims to solve this issue by learning representations from the spectral components of CSI features, which can more effectively help bridge the domain gap. Accordingly, we propose a spectral-attentive framework, named SCA-LLM, for channel prediction in multiple-input multiple-output orthogonal frequency division multiplexing (MIMO-OFDM) systems. Specifically, its novel adapter can capture finer spectral details and better adapt the LLM for channel prediction than previous methods. Extensive simulations show that SCA-LLM achieves state-of-the-art prediction performance and strong generalization, yielding up to $-2.4~\text{dB}$ normalized mean squared error (NMSE) advantage over the previous LLM based method. Ablation studies further confirm the superiority of SCA-LLM in mitigating domain mismatch.
\end{abstract}


\section{Introduction}
Artificial intelligence-radio access network (AI-RAN) is transforming wireless system design from hand-crafted model-driven optimization toward data-driven agentic communications \cite{kundu2025ai, DBLP:6G}. In this paradigm, the network is expected to behave as an agentic system that can continuously sense, predict, and perform look-ahead planning under fast-varying propagation conditions, rather than merely reacting to the instantaneous channel state \cite{HeRisk2024, HeXoA2025}. For modern multiple-input multiple-output orthogonal frequency division multiplexing (MIMO-OFDM) deployments \cite{DBLP:MIMO-OFDM,DBLP:MIMO1,DBLP:OFDM}, adaptive techniques such as dynamic resource allocation \cite{DBLP:resource_allocation}, antenna selection \cite{HeRisk2024}, precoding \cite{DBLP:precoding}, and beamforming \cite{DBLP:beamforming1} are increasingly time-sensitive and uncertainty-aware. However, practical CSI acquisition and feedback inevitably incur latency and overhead \cite{HeXoA2025}, and the resulting channel aging makes instantaneous CSI insufficient to reliably guide uncertainty-aware optimizations in dynamic environments \cite{DBLP:Channel_Aging}. Therefore, the ability to forecast the multi-step evolution of channel state information (CSI) becomes a key enabler for reliable and efficient look-ahead decision-making in agentic communications \cite{HeRisk2024, HeTowards2022}.

From this perspective, multi-step CSI prediction naturally serves as a short-horizon \emph{wireless world model} that predicts the transition dynamics of the propagation environment. By rolling out future CSI trajectories, the network agent gains the ability to reason about the near-future channel evolution and to proactively plan transmission actions, rather than reacting to a possibly stale CSI snapshot \cite{HeRisk2024}. Recognizing this, recent research has begun to leverage wireless world models for agentic communications. For example, the authors in \cite{HeRisk2024} and \cite{HeXoA2025} proposed to use deep generative models as the wireless world model, and perform risk-aware antenna selection via look-ahead planning on the near-future CSI evolutions. 

Despite these encouraging results, existing wireless world models are typically small-scale and built upon scenario-specific models \cite{HeRisk2024, HeXoA2025, DBLP:PAD, DBLP:ST-AR, DBLP:cp_kf_vs_mL, DBLP:RNN, DBLP:fading_cp, DBLP:LSTM, DBLP:GRU, DBLP:Transformer}, whose generalization is often limited when facing heterogeneous deployments, varying mobility patterns, and diverse radio configurations. As AI-RAN scales up and operates in continually changing environments, we increasingly need a reusable foundational sequence-model backbone for wireless world modeling, which should encode strong dynamics priors, transfer well across scenarios, and admit parameter-efficient adaptation. This is also the main focus of this work.

% Recent progress in large language models (LLMs) suggests a promising direction: although originally developed for text, LLMs are powerful general-purpose sequence learners with strong capacity to model long-range dependencies, making them attractive as universal dynamics backbones when appropriately adapted to non-text time-series signals. However, bridging CSI to LLMs is non-trivial. CSI is a structured complex-valued spatio-temporal-frequency tensor with domain-specific spectral characteristics, and a naive tokenization or generic pooling-based adapter may destroy the very inductive biases needed for modeling channel dynamics. Therefore, an effective adapter must expose informative spectral components and temporal evolution patterns to align CSI with the LLM’s sequence modeling capability, while maintaining parameter efficiency and robustness to domain mismatch.

\subsection{Prior Works}
Existing channel prediction methods can be broadly categorized into traditional methods and deep learning (DL) based methods.
Traditional methods are typically based on explicit mathematical models derived from statistical assumptions or simplified representations of the underlying channel physics.
However, it is often difficult for these methods to perform well in complex and dynamic wireless environments.
For instance, although advanced model-based techniques, such as the Prony-based angle-delay domain (PAD) method \cite{DBLP:PAD}, are designed to capture specific channel structures and potentially offer improved accuracy under certain conditions, fundamental limitations are often encountered.
While analytical guarantees are provided by PAD, frequent updates to its model parameters are required for adaptation to environmental changes\cite{DBLP:ST-AR}. Substantial computational overhead is incurred by this process as well.
There is still a need for more robust and generalizable channel prediction approaches capable of effective adaptation to diverse environments.

In response to this need, DL based methods \cite{DBLP:ST-AR, DBLP:cp_kf_vs_mL, DBLP:RNN, DBLP:fading_cp, DBLP:LSTM, DBLP:GRU, DBLP:Transformer} have emerged as a promising alternative. 
Rather than relying on predefined mathematical models, DL methods are data-driven approaches that learn complex dependencies and predictive patterns directly from historical channel observations.
Architectures such as recurrent neural networks (RNNs) \cite{DBLP:RNN}, long short-term memory (LSTM) networks \cite{DBLP:LSTM}, gated recurrent units (GRUs) \cite{DBLP:GRU}, and transformers \cite{DBLP:Transformer} have been employed, attempting to model the complex spatio-temporal dynamics of wireless channels. 
For example, focusing on the angle-delay domain for high-mobility massive MIMO systems, a complex-valued neural network (CVNN) based prediction method has been introduced in \cite{DBLP:ST-AR}. 
This data-driven approach performs element-wise prediction of the angle-delay channel response matrix (ADCRM). 
By learning directly from historical channel data represented in the angle-delay domain, this method achieved improved prediction accuracy compared to the traditional model-based PAD method \cite{DBLP:PAD}.
Furthermore, transformer-based methods for channel prediction have been introduced in \cite{HeXoA2025} and \cite{DBLP:Transformer}. These approaches use the attention mechanism within the transformer model to establish the relationship between past CSI and future channels.
It gives more importance to the most relevant past information, which improves the prediction accuracy and overall communication performance.
Additionally, a unified pilot-to-precoder scheme has been proposed either in \cite{DBLP:Transformer}, using a single transformer model to integrate channel estimation, prediction, and precoding, thereby reducing computational complexity.
While these approaches have often demonstrated improved accuracy over traditional methods, they still have limitations in practice. 
Specifically, these models are typically trained for specific tasks and environments. 
When encountering unseen communication environments without significant retraining, they often exhibit a lack of robustness due to limited sequence modeling capabilities.
Therefore, there is a need to develop novel approaches that are more robust for unseen environments.

In recent years, large language models (LLMs) \cite{DBLP:Qwen2, DBLP:Llama3, DBLP:DeepSeek-R1, GPT-2} have shown impressive abilities in pattern recognition and sequence modeling thanks to training on massive amounts of sequential data. 
Their success has led to a growing interest in using LLMs to solve problems in the physical layer of wireless communications \cite{DBLP:WirelessLLM, DBLP:LLM4TeleCom, DBLP:WirelessGPT, DBLP:NetGPT}. 
One area of particular interest is channel prediction, which naturally fits the strength of LLMs as a sequential forecasting task. 
However, directly applying LLMs, which are pre-trained primarily on textual data to specialized physical layer signals such as wireless channels, often leads to a significant \textit{domain mismatch}. 
LLMs, pre-trained mainly on text data, lack a good understanding of the complex spatio-temporal dynamics governed by wireless channel characteristics.
Therefore, directly applying LLMs to raw or minimally processed CSI data significantly restricts their predictive capabilities for channel prediction tasks, which will be shown by the ablation study presented later in this work. 

Recognizing this, the recent works such as LLM4CP \cite{DBLP:LLM4CP} have proposed the ``adapter + LLM'' paradigm for channel prediction, where a specialized adapter preprocesses the CSI data before feeding them to a pre-trained LLM.
The adapter of LLM4CP employs channel attention layers inspired by the SE-Net \cite{DBLP:SENet}, termed CSI-attention modules, for feature analysis and modeling. 
These modules attempt to capture the spatio-temporal correlations within the CSI data and transform the data into a feature representation more easily interpreted by the subsequent LLM, aiming to mitigate the domain mismatch issue.
As an effective initial effort, LLM4CP validated the feasibility and potential of the ``adapter + LLM'' paradigm. 
The effectiveness of the LLM4CP's adapter relies on the global average pooling (GAP) operation within its CSI-attention modules to generate the channel attention weights. 
By using GAP, only the lowest frequency component of the spatio-temporal feature spectrum is preserved, while all other frequency components are discarded \cite{DBLP:FcaNet}. 
Although preserving only the lowest frequency component, experiments in \cite{DBLP:LLM4CP} suggested this representation learning method still offers some benefit for modeling multipath effect and extracting underlying physical propagation features.

However, the discarded spectral components also encode valuable information about the underlying wireless channel characteristics. 
This information reflects phenomena such as multipath effect within specific scattering environments, spatio-temporal correlations across the antenna array, mobility-induced Doppler effects, and environmental noise during propagation.
Since accurate modeling of MIMO channel dynamics relies on such details, discarding them consequently restricts the performance potential of LLM4CP. 
Therefore, despite these early advances, there remains a significant gap for existing methods to fully exploiting LLMs for channel prediction. 

\subsection{Contributions}
To fill the gap, this work propose SCA-LLM, a spectral-attentive framework with a LLM well pre-trained on a very large corpus of general knowledge data, for accurate channel prediction in MIMO-OFDM systems.
The proposed SCA-LLM comprises three sequential components: the spectral-attentive adapter, a pre-trained LLM, and the output head. 
% It is necessary to clarify that the term \textit{spectral-attentive} here refers to the analysis of the spectral components within features transformed from the CSI data.
% The adapter consists sequentially of the CSI embedding layer and the so-called spectral channel adaptation (SCA) module.
% For the pre-trained LLM, we use the same selective fine-tuning strategy as used in LLM4CP to preserve the vast majority of the LLM’s pre-trained general knowledge, where only the layer normalization and positional embedding layers are fine-tuned.
Specifically, the main contributions of this work are summarized as follows. 

\begin{enumerate}
    \item We propose the SCA-LLM, a novel spectral-attentive framework for MIMO-OFDM channel prediction, which integrates a pre-trained LLM with a specialized spectral-attentive adapter to more effectively mitigate the domain-mismatch issue.
    \item We provide a key insight that learning representations from the spectral components of CSI features can more effectively help bridge the gap between the CSI data and the LLM. 
    Accordingly, we introduce the spectral channel adaptation (SCA) module within the adapter. 
    This module adopts multi-spectral channel attention layers, which utilize the 2-dimensional (2-D) discrete cosine transform (DCT) \cite{DBLP:JPEG_DCT, DBLP:FcaNet} for feature analysis and modeling. 
    Unlike LLM4CP's CSI-attention modules, the SCA module leverages multiple DCT bases to selectively preserve multiple spectral components within features, ranging from low frequency components to high frequency components. 
    This helps the SCA-LLM to better model the wireless channel characteristics and bridge the domain gap between the CSI data and the LLM more effectively, thereby improving the prediction performance significantly.
    % The SCA module utilizes multi-spectral channel attention layers based on the 2-D DCT to preserve a richer useful set of spectral components from CSI-derived features compared to previous approaches.
    \item Extensive simulations demonstrate that the proposed SCA-LLM achieves state-of-the-art (SOTA) prediction performance and exhibits strong robustness and \emph{zero-shot} generalization capabilities, across diverse communication scenarios, mobility conditions, and signal-to-noise ratios (SNRs). Ablation studies further validate the effectiveness and superiority of the proposed SCA-LLM, confirming its significant contribution to adapting the LLM for channel prediction task.
\end{enumerate}

\subsection{Organization and Notation}

\subsubsection{\textit{Organization}}
The remainder of this paper is organized as follows. Section II introduces the system model. Section III formulates the channel prediction problem. Section IV details the proposed SCA-LLM framework. Section V presents the numerical results and discusses the performance evaluation. Finally, Section VI concludes the paper.

\subsubsection{\textit{Notation}}
Throughout this paper, we use bold uppercase letters, e.g., $\mathbf{H}$, to denote matrices and bold lowercase letters, e.g., $\mathbf{a}$, for vectors. The operators $(\cdot)^T$, $(\cdot)^*$, and $(\cdot)^H$ represent the transpose, complex conjugate, and hermitian transpose (conjugate transpose), respectively. The operators $\text{Re}(\cdot)$ and $\text{Im}(\cdot)$ extract the real and imaginary parts of a complex number or matrix, respectively. The $\text{vec}(\cdot)$ operator vectorizes a matrix by stacking its columns into a single column vector. The squared Frobenius norm operator is denoted by $||\cdot||_F^2$. ZOA, ZOD, AOA, and AOD stand for zenith angle of arrival, zenith angle of departure, azimuth angle of arrival, and azimuth angle of departure, respectively.

\section{System Model}
\subsection{MIMO-OFDM System}
In this work, we consider a MIMO-OFDM communication system where a base station (BS) equipped with $N_{BS}$ antennas communicates with user equipments (UEs) with $N_{UE}$ antennas. In order to optimize the data transmission to the users, the BS must conduct CSI estimation for all users. Assuming orthogonal pilot sequences across the users, the CSI estimation and prediction procedure can be applied to every user independently. Without loss of generality, we focus on a particular active user in the system, thus omit the user index \cite{DBLP:Transformer, DBLP:journals/tcom/YangGZAA20}. The total bandwidth $B$ is divided into $K$ orthogonal subcarriers, indexed by $k=0, 1, \dots, K-1$, with subcarrier spacing $\Delta f$. 
A cyclic prefix (CP), configured to be longer than the maximum channel delay spread, is added to each OFDM symbol. 
Moreover, the system operates in time-division duplexing (TDD) mode, where the uplink (UL) and downlink (DL) transmissions use the same frequency band but occur at different time slots within a communication frame. 
Each frame allocates specific time slots for transmitting uplink pilot signals, such as sounding reference signals (SRS) from the UE.
These pilots enable the BS to periodically estimate the UL channel state information (CSI).
Following the pilot transmissions, the frame structure provides configurable time slots designated for UL and DL data transmission.

\subsection{Channel Model}
To practically reflect the wireless propagation environment, this work adopts the channel model based on the 3GPP TR 38.901 standard \cite{3gpp_tr38901_2018}. 
This channel model is the implementation of the geometry-based stochastic channel model (GSCM) framework \cite{DBLP:GSCM_overview}.
Specifically, the UL channel between the BS and UE can be modeled as:
\begin{equation}
\begin{split}
    \mathbf{H}_{UL}(t, f) ={}& \sum_{n=1}^{N} \sum_{m=1}^{M} \alpha_{n,m} e^{-j 2\pi f \tau_{n,m}} e^{j 2\pi \nu_{n,m} t}  \mathbf{a}_{BS}(\theta_{n,m, ZOA}, \phi_{n,m, AOA}) \mathbf{a}_{UE}^H(\theta_{n,m, ZOD}, \phi_{n,m, AOD}), % 在 UE 向量前增加更多缩进 \qquad
\end{split}
\label{eq:channel_model_formatted} % 可以修改标签以示区别
\end{equation}
where $N$ is the number of propagation clusters, and $M$ is the number of rays within each cluster. 
For the $m$-th ray of the $n$-th cluster, $\alpha_{n,m}$ is its complex gain, $\tau_{n,m}$ is its delay, and $\nu_{n,m}$ is its Doppler shift. 
$f$ denotes the frequency of the given carrier, and $t$ represents the given time. 
The vectors $\mathbf{a}_{BS}(\theta_{n,m, ZOA}, \phi_{n,m, AOA})$ and $\mathbf{a}_{UE}(\theta_{n,m, ZOD}, \phi_{n,m, AOD})$ are the array response vectors of the BS and UE antennas, respectively. 
These two vectors depend on the ray's ZOA ($\theta_{n,m, ZOA}$), AOA ($\phi_{n,m, AOA}$), ZOD ($\theta_{n,m, ZOD}$), and AOD ($\phi_{n,m, AOD}$). 
The Doppler frequency component $\nu_{n,m}$ depends on the arrival angles ($\theta_{n,m, ZOA}, \phi_{n,m, AOA}$) and the UE velocity vector ${\mathbf{v}}$.

The above channel model can account for factors like multipath propagation through scattering clusters, path delays, and the angular spread of signals.
This indicates that the channel model can capture dynamic spatio-temporal correlations of the MIMO-OFDM channel, which is influenced by factors such as user mobility and variations in the scattering environment.
In addition, the specific parameters within the channel model, which dictate channel characteristics, are determined by the chosen deployment scenario, such as urban macro (UMa) or urban micro (UMi).

\section{Problem Formulation of Channel Prediction}
In this work, our goal is to predict the future DL CSI sequence based on the historical UL CSI sequence.
Predicting future DL CSI allows the base station to prepare efficient transmission strategies, such as beamforming, ahead of time. 
This avoids the delays and overhead associated with measuring the current channel state, leading to faster and more reliable communication. 

\subsection{TDD Operation and Channel Reciprocity}
Since the system adopts the TDD mode, we can leverage the channel reciprocity. 
This property indicates that the DL CSI at a certain time $t$ and a subcarrier $k$ with frequency $f$, denoted as $\mathbf{H}_{DL}(t, f) \in \mathbb{C}^{N_{UE} \times N_{BS}}$, can be directly obtained from the corresponding UL CSI $\mathbf{H}_{UL}(t, f) \in \mathbb{C}^{N_{BS} \times N_{UE}}$ by:
\begin{equation} \label{eq:reciprocity}
    \mathbf{H}_{\mathrm{DL}} = \mathbf{H}_{\mathrm{UL}}^{T}.
\end{equation}
Leveraging this reciprocity, the channel prediction problem is simplified from DL channel acquisition complexities, and predicting future UL CSI becomes sufficient to determine future DL CSI.
This simplification therefore makes it much easier to directly evaluate our method's prediction performance and clearly demonstrate its core contribution.

\subsection{Channel Prediction Problem Formulation}
Based on the channel reciprocity discussed above, our task simplifies to predicting future UL CSI using historical observations. 
This historical CSI sequence is typically acquired through channel estimation at previous time steps, or directly incorporating previously predicted CSI samples. 
For the estimation process, the UE transmits known pilot sequences, denoted by $\mathbf{X}_p \in \mathbb{C}^{N_{UE} \times T_p}$, over $T_p$ symbol durations within designated time slots. 
The corresponding signal received at the BS, $\mathbf{Y}_p \in \mathbb{C}^{N_{BS} \times T_p}$, can be modeled as:
\begin{equation} \label{eq:channel_estimation_model}
    \mathbf{Y}_p(t) = \mathbf{H}_{\mathrm{UL}}(t, f) \mathbf{X}_p + \mathbf{N}(t),
\end{equation}
where $\mathbf{H}_{\mathrm{UL}}(t, f) \in \mathbb{C}^{N_{BS} \times N_{UE}}$ is the UL channel matrix at time $t$ and frequency $f$, and $\mathbf{N}(t) \in \mathbb{C}^{N_{BS} \times T_p}$ represents the additive noise. 
For notational clarity, we often omit the frequency index $f$ from $\mathbf{H}_{\mathrm{UL}}(t, f)$ and use $\mathbf{H}_{\mathrm{UL}}(t)$, assuming the operations are performed per frequency, or that the model implicitly handles the frequency dimension. 
Channel estimation techniques, such as least squares (LS) or minimum mean square error (MMSE), can then be applied to obtain an estimate $\hat{\mathbf{H}}_{\mathrm{UL}}(t)$ from $\mathbf{Y}_p(t)$ and the known $\mathbf{X}_p$.

These channel estimates, possibly augmented with prior predictions, are from the historical data required for the current channel prediction task. 
Let the sequence of available UL CSI samples over a past window of $L$ time steps ending at the current time $t_{0}$ be denoted by:
\begin{equation} \label{eq:past_csi}
    \mathcal{H}_{\mathrm{past}} = \{\mathbf{H}_{\mathrm{UL}}(t_{0}-L+1), \dots, \mathbf{H}_{\mathrm{UL}}(t_{0}-1), \mathbf{H}_{\mathrm{UL}}(t_{0}) \},
\end{equation}
where each $\mathbf{H}_{\mathrm{UL}}(t)$ in the sequence represents the most recently available CSI for that time step, obtained via estimation or prediction. 

The objective of this work is to predict the sequence of UL CSI samples for the next $P$ future time steps, starting from $t_0+1$:
\begin{equation} \label{eq:future_csi}
    \mathcal{H}_{\mathrm{future}} = \{\mathbf{H}_{\mathrm{UL}}(t_{0}+1), \mathbf{H}_{\mathrm{UL}}(t_{0}+2), \dots, \mathbf{H}_{\mathrm{UL}}(t_{0}+P)\}.
\end{equation}
We aim to learn a predictive model, represented by a function $\mathcal{F}$, that maps the historical sequence $\mathcal{H}_{\mathrm{past}}$ to the future sequence $\mathcal{H}_{\mathrm{future}}$:
\begin{equation} \label{eq:prediction_func}
    \hat{\mathcal{H}}_{\mathrm{future}} = \mathcal{F}(\mathcal{H}_{\mathrm{past}}; \Theta),
\end{equation}
where $\hat{\mathcal{H}}_{\mathrm{future}} = \{\hat{\mathbf{H}}_{\mathrm{UL}}(t_{0}+1), \hat{\mathbf{H}}_{\mathrm{UL}}(t_{0}+2), \dots, \hat{\mathbf{H}}_{\mathrm{UL}}(t_{0}+P)\}$ is the sequence of predicted future UL CSI samples, and $\Theta$ represents the learnable parameters of the prediction model $\mathcal{F}$. 
The model $\mathcal{F}$ is typically trained by minimizing a loss function that quantifies the discrepancy between the predicted sequence $\hat{\mathcal{H}}_{\mathrm{future}}$ and the ground-truth sequence $\mathcal{H}_{\mathrm{future}}$, such as the mean squared error (MSE) or the normalized mean squared error (NMSE).

Once the future UL CSI sequence $\hat{\mathcal{H}}_{\mathrm{future}}$ is predicted, the corresponding future DL CSI sequence $\hat{\mathcal{H}}_{\mathrm{DL, future}} = \{\hat{\mathbf{H}}_{\mathrm{DL}}(t_{0}+1), \dots, \hat{\mathbf{H}}_{\mathrm{DL}}(t_{0}+P)\}$ can be readily obtained via channel reciprocity using (\ref{eq:reciprocity}):
\begin{equation} \label{eq:dl_prediction}
    \hat{\mathbf{H}}_{\mathrm{DL}}(t_{0}+p) = \left(\hat{\mathbf{H}}_{\mathrm{UL}}(t_{0}+p)\right)^T, \quad \text{for } p = 1, \dots, P.
\end{equation}

Therefore, the channel prediction problem is defined as a sequence forecasting task where the input is a multivariate time series of past MIMO CSI samples, and the output is the predicted sequence of future MIMO CSI samples. These time series essentially reflect the complex spatio-temporal correlations within the MIMO channel, which arises from factors such as user mobility and the scattering environment. Accurately leveraging these correlations for prediction is the core challenge that needs to be addressed by the prediction model $\mathcal{F}$.

\section{The Proposed SCA-LLM Framework}
\label{sec:proposed_framework}

\begin{figure*}[!htbp]
    \centering
    \includegraphics[width=0.8\linewidth]{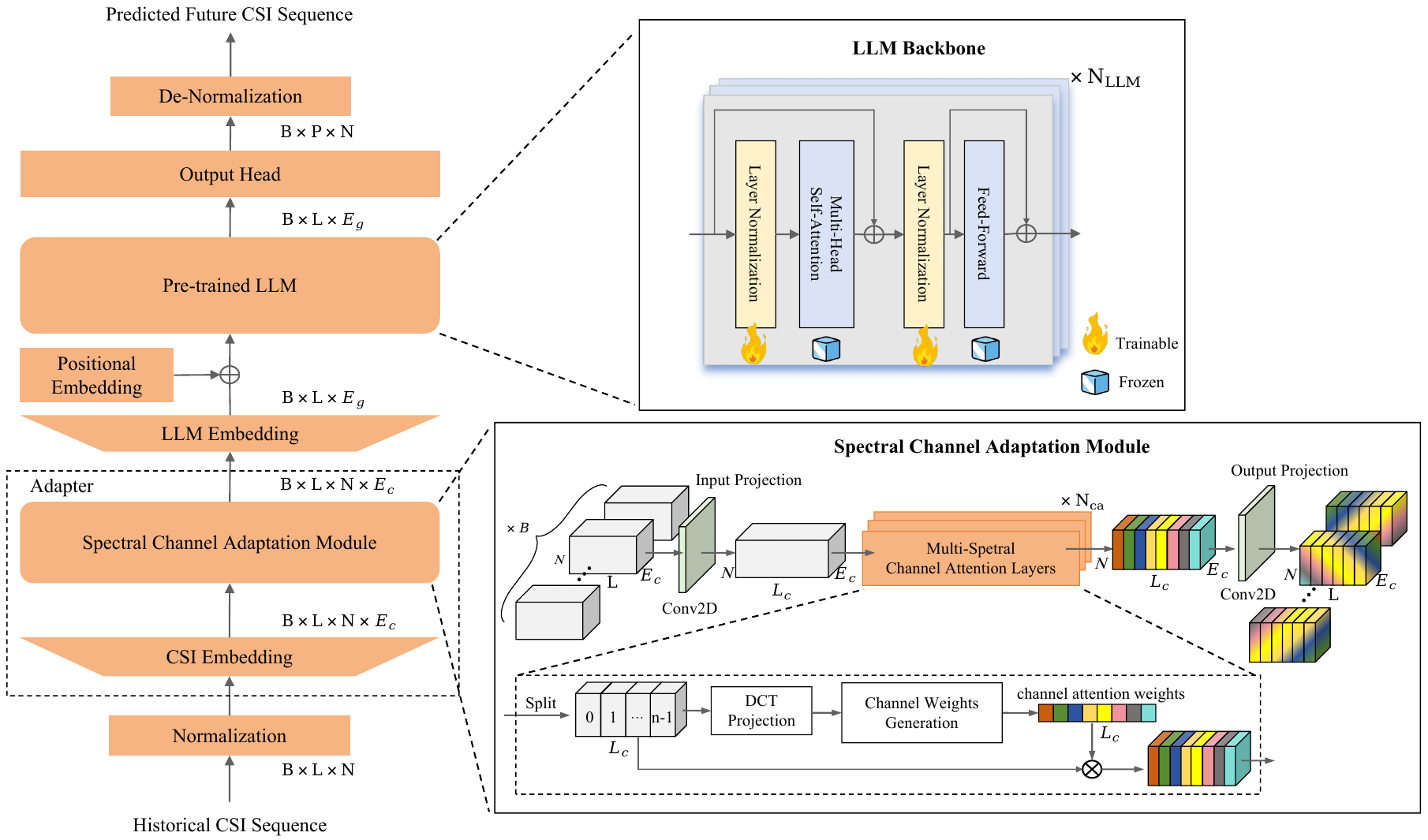}
    \caption{The architecture of the proposed SCA-LLM framework. The adapter component encompasses the CSI embedding layer and the spectral channel adaptation (SCA) module.}
    \label{fig:model_architecture}
\end{figure*}

In this section, we introduce the proposed SCA-LLM framework, which is designed to effectively adapt pre-trained LLMs for channel prediction. 
The overall architecture of the SCA-LLM is shown in Fig.~\ref{fig:model_architecture}. 
This framework processes the historical CSI sequence sequentially through the following components: 1) the normalization layer; 2) the spectral-attentive adapter consisting of the CSI embedding layer and the spectral channel adaptation (SCA) module; 3) the pre-trained LLM consisting of the LLM embedding layer with positional embedding and the LLM backbone; 4) the output head; 5) and finally the de-normalization layer.

Since the SCA-LLM framework operates on real-valued inputs, the complex-valued CSI data should be converted into a suitable real-valued representation. 
To this end, the complex-valued CSI matrix $\mathbf{H}_{\mathrm{UL}}(t) \in \mathbb{C}^{N_{BS} \times N_{UE}}$ for each subcarrier $k$ at time step $t$ is first vectorized and converted into a real-valued representation $\mathbf{h}(t) \in \mathbb{R}^{N}$, which is given by:
\begin{equation} \label{eq:vectorization}
    \mathbf{h}(t) = \begin{bmatrix} \text{Re}(\text{vec}(\mathbf{H}_{\mathrm{UL}}(t))) \\ \text{Im}(\text{vec}(\mathbf{H}_{\mathrm{UL}}(t))) \end{bmatrix},
\end{equation}
where $N = 2 N_{BS} N_{UE}$. 
After that, the historical CSI sequence is formed by collecting these vectors $\mathbf{h}(t)$ over $L$ time steps, denoted as $\mathbf{H} \in \mathbb{R}^{B \times L \times N}$, where $B$ is the batch size and $L$ is the number of past time steps.
Additionally, since the subcarriers within OFDM are orthogonal, we can process each subcarrier in parallel \cite{DBLP:MIMO-OFDM} for computational efficiency.
That is, we rearrange the CSI data such that the individual subcarriers within the historical CSI sequence are merged into the batch dimension $B$.
After the above operations, the historical CSI sequence $\mathbf{H} \in \mathbb{R}^{B \times L \times N}$ is prepared to serve as the input to the SCA-LLM and is processed sequentially by the following components.

\subsection{Normalization}
This step aims to standardize the input data distribution, which often helps stabilize model training and improve convergence speed. 
The normalization uses the mean and standard deviation calculated across all elements within the input tensor $\mathbf{H} \in \mathbb{R}^{B \times L \times N}$, which is expressed as:
\begin{equation} \label{eq:norm_layer}
    \mathbf{H}_{\text{norm}} = \frac{\mathbf{H} - \mu_{\mathbf{H}}}{\sigma_{\mathbf{H}}},
\end{equation}
where $\mu_{\mathbf{H}}$ and $\sigma_{\mathbf{H}}$ denote the mean and standard deviation computed over all dimensions (batch, time, and feature) of the input tensor $\mathbf{H}$. 
The resulting normalized tensor $\mathbf{H}_{\text{norm}}$ serves as the input to the subsequent spectral-attentive adapter.
In addition, these calculated statistics, $\mu_{\mathbf{H}}$ and $\sigma_{\mathbf{H}}$, are stored and subsequently used by the de-normalization layer to transform the model's final predictions back into the original scale of the input CSI data.

\subsection{Adapter}

\subsubsection{CSI Embedding}
The CSI embedding layer receives the normalized historical CSI sequence, $\mathbf{H}_{\text{norm}} \in  \mathbb{R}^{B\times L \times N}$, as the input. 
This layer transforms the $N$-dimensional ($N\text{-D}$) feature vector at each time step (corresponding to the last dimension of $\mathbf{H}_{\text{norm}}$) into an embedding representation within a higher-dimensional latent space, resulting in a $4\text{-D}$ tensor $\mathbf{H}_{\text{emb}} \in \mathbb{R}^{B \times L \times N \times E_{c}}$ as the output.
Here, $E_{c}$ is the target embedding dimension for each of the original $N$ features.

Specifically, this transformation is achieved through two sequential point-wise linear projections, which operate independently on the feature vector at each time step. Mathematically, this process can be formulated as:
\begin{equation} \label{eq:csi_embedding}
    \mathbf{H}_{\text{emb}} = \text{Reshape}_{B,L,N,E_c} \left( \text{Linear}_2 \left( \text{Linear}_1 \left( \mathbf{H}_{\text{norm}} \right) \right) \right),
\end{equation}
where $\text{Linear}_1$ represents the first point-wise linear projection (mapping $N \rightarrow N_{hs}$ features) and $\text{Linear}_2$ represents the second point-wise linear projection (mapping $N_{hs} \rightarrow N \cdot E_c$ features). 
The $\text{Reshape}_{B,L,N,E_c}(\cdot)$ function rearranges the last dimension of the output tensor from $(N \cdot E_c)$ into $(N, E_c)$.
In practice, these projections are implemented using $1\text{-D}$ convolutional layers with a kernel size of $1$ and a stride of $1$ for computational efficiency.
This transformation is designed to prepare the data for the subsequent SCA module, and the embedded representation $\mathbf{H}_{\text{emb}}$ serves as the input to the subsequent SCA module.

\subsubsection{Spectral Channel Adaptation Module}
This module is the core component designed to bridge the domain gap between the CSI data and the pre-trained LLM. 
As highlighted in the Introduction, directly applying text-pretrained LLMs to CSI data faces challenges due to domain mismatch. 
The SCA module addresses this by adaptively refining the features based on their spectral components. 
Specifically, it employs multi-spectral channel attention layers that utilize the 2-D discrete cosine transform (DCT) to capture richer information than methods relying solely on the lowest frequency component (like GAP used in \cite{DBLP:LLM4CP}). 
This allows the module to better model wireless channel characteristics and adapt the features into a format more effectively processed by the subsequent LLM. 
The architecture of this module, detailed in the lower panel of Fig.~\ref{fig:model_architecture}, consists of an input projection layer, a stack of multi-spectral channel attention layers, and an output projection layer.

\begin{itemize}
    \item \textbf{Input Projection}: 
        The first step within the SCA module is an input projection layer. 
        It receives the output from the CSI embedding layer, $\mathbf{H}_{\text{emb}} \in \mathbb{R}^{B \times L \times N \times E_{c}}$. 
        Its purpose is to transform the input features while mapping the temporal dimension $L$ to an intermediate dimension $L_c$, suitable for the subsequent multi-spectral channel attention layers. 
        This transformation is achieved using a 2-D convolutional layer. 
        Specifically, the convolution operates on the $(N, E_c)$ spatial dimensions with a kernel size of $(3, 3)$ and appropriate padding to maintain these dimensions.
        This operation takes the $L$ feature maps corresponding to the time dimension as input and produces $L_c$ output feature maps.
        Mathematically, this can be expressed as:
        \begin{equation} \label{eq:sca_input_proj_conv3x3}
             \mathbf{H}_{c}^{(0)} = \text{Conv2D}_{k=(3,3)}^{(L \to L_c)} \left( \mathbf{H}_{\text{emb}} \right), 
        \end{equation}
        where $\text{Conv2D}_{k=(3,3)}^{(L \to L_c)}(\cdot)$ denotes the 2-D convolution operation described above. 
        This operation mixes local spatial information within the $(N, E_c)$ plane while projecting the time dimension $L$ to $L_{c}$, resulting in $\mathbf{H}_{c}^{(0)} \in \mathbb{R}^{B \times L_c \times N \times E_{c}}$. 
        This tensor serves as the input to the first multi-spectral channel attention layer.

    \item \textbf{Multi-Spectral Channel Attention Layers}:
        After that, a stack of $N_{ca}$ multi-spectral channel attention layers \cite{DBLP:FcaNet}, are applied sequentially to these feature maps for feature analysis and modeling. Let $\mathbf{H}_{c}^{(i)}$ be the input of the $i\text{-th}$ multi-spectral channel attention layer ($\mathbf{H}_{c}^{(0)} = \mathbf{H}_{c}$) and $MSCA(\cdot)$ denote the operator of each layer, then we have:
        \begin{equation}
            \mathbf{H}_{c}^{(i+1)} = MSCA(\mathbf{h}_{c}^{(i)}) \in \mathbb{R}^{B \times L_{c} \times N \times E_{c}}.
        \end{equation}
        For improved readability and to avoid excessive detail in this section, a full description of the multi-spectral channel attention layer is available in Appendix~\ref{layer: msca}.

    \item \textbf{Output Projection}:
    Following the multi-spectral channel attention layers, the resulting tensor $\mathbf{H}_{c}^{(N_{ca})} \in \mathbb{R}^{B \times L_c \times N \times E_{c}}$ is processed by the output projection layer. 
    This layer maps the intermediate dimension $L_c$ back to the original time dimension $L$. 
    Similar to the input projection, this transformation is implemented using a 2-D convolutional layer, mapping $L_c$ feature maps as the input to $L$ feature maps as the output. 
    Mathematically, this process is given by:
    \begin{equation} \label{eq:sca_output_proj_conv3x3}
        \mathbf{H}_{\text{sca}} = \text{Conv2D}_{k=(3,3)}^{(L_c \to L)} \left( \mathbf{H}_{c}^{(N_{ca})} \right).
    \end{equation}
    The resulting tensor $\mathbf{H}_{\text{sca}} \in \mathbb{R}^{B \times L \times N \times E_{c}}$ is the final output of the SCA module, which is then passed to the LLM embedding layer.

\end{itemize}

\subsection{Pre-trained LLM}
\subsubsection{LLM Embedding}
Following the SCA module, the resulting tensor, let's denote it as $\mathbf{H}_{\text{sca}} \in \mathbb{R}^{B \times L \times N \times E_{c}}$, is fed into the LLM embedding layer. 
The purpose of this layer is to project the adapted CSI features into the embedding space expected by the pre-trained LLM, which has a dimension of $E_g$.

Specifically, the transformation first involves flattening the last two dimensions ($N$ and $E_c$) of $\mathbf{H}_{\text{sca}}$. 
Then, a linear projection is applied to map the resulting $N \cdot E_c$-dimensional features to the target $E_g$ dimension. 
Mathematically, this projection step can be represented as:
\begin{equation} \label{eq:llm_embedding_proj}
    \mathbf{H}_{\text{LLM\_emb}} = \text{Linear}_{\text{Emb}} \left( \text{Flatten}_{N,E_c} \left( \mathbf{H}_{\text{sca}} \right) \right),
\end{equation}
where $\text{Flatten}_{N,E_c}(\cdot)$ denotes the operation that combines the last two dimensions, and $\text{Linear}_{\text{Emb}}$ represents the linear projection layer mapping $N \cdot E_c \rightarrow E_g$ features. The output tensor $\mathbf{H}_{\text{LLM\_emb}} \in \mathbb{R}^{B \times L \times E_g}$. 
In practice, $\text{Linear}_{\text{Emb}}$ is efficiently implemented using a $1\text{-D}$ convolutional layer too.

Furthermore, to provide the LLM with information about the order of time steps within the sequence, positional embeddings $\mathbf{P} \in \mathbb{R}^{L \times E_g}$ are added to the projected features. 
The specific method for generating $\mathbf{P}$ (e.g., learned embeddings, sinusoidal functions) depends on the chosen LLM architecture. 
The addition operation is broadcast across the batch dimension $B$:
\begin{equation} \label{eq:llm_embedding_pos}
    \mathbf{X}_{\text{in}} = \mathbf{H}_{\text{LLM\_emb}} + \mathbf{P}.
\end{equation}
The resulting tensor $\mathbf{X}_{\text{in}} \in \mathbb{R}^{B \times L \times E_g}$ serves as the final input sequence to the LLM backbone.

\subsubsection{LLM Backbone}
The core sequence processing is performed by the pre-trained LLM backbone. 
In this work, we utilize GPT-2 \cite{GPT-2} as the foundational LLM. 
The LLM takes the prepared input sequence $\mathbf{X}_{\text{in}} \in \mathbb{R}^{B \times L \times E_g}$ and processes it through its stack of transformer layers, denoted collectively as $LLM(\cdot)$. 
This transformation yields the final hidden state sequence:
\begin{equation} \label{eq:llm_backbone}
    \mathbf{H}_{\text{LLM}} = LLM(\mathbf{X}_{\text{in}}),
\end{equation}
where $\mathbf{H}_{\text{LLM}} \in \mathbb{R}^{B \times L \times E_g}$ represents the sequence of output hidden states from the LLM.

As depicted in the upper panel of Fig.~\ref{fig:model_architecture}, the GPT-2 architecture consists of $N_{\text{LLM}}$ stacked transformer decoder blocks built upon standard components: multi-head self-attention (MHSA), layer normalization (LN), and feed-forward networks (FFN). 
To effectively adapt the pre-trained model for the channel prediction task while preserving its powerful general sequence modeling capabilities acquired during large-scale pre-training, we employ a selective fine-tuning strategy. 
As indicated by the symbols in Fig.~\ref{fig:model_architecture}, the parameters of the MHSA and FFN layers are kept frozen (ice cube symbol), retaining the LLM's core knowledge. 
Conversely, the parameters within the LN layers are fine-tuned (flame symbol), allowing the model to adapt specifically to the statistical properties and nuances of the embedded CSI data. 
This strategy aims to bridge the domain gap effectively while leveraging the LLM's pre-trained strengths.
Notably, the proposed SCA-LLM framework is flexible regarding the choice of the LLM backbone. 
Alternatives such as Qwen2 \cite{DBLP:Qwen2} or Llama3 \cite{DBLP:Llama3} could potentially be integrated. 
The selection depends on the specific requirements concerning computational resources and desired prediction performance.

\subsection{Output Head}
The output head module receives the final hidden state sequence from the pre-trained LLM, denoted as $\mathbf{H}_{\text{LLM}} \in \mathbb{R}^{B \times L \times E_g}$. 
Its purpose is to transform this sequence into the final normalized prediction sequence $\hat{\mathbf{H}}_{\text{norm}} \in \mathbb{R}^{B \times P \times N}$, where $P$ is the desired future prediction length. 
This transformation involves two main steps: mapping the feature dimension from $E_g$ back to $N$, and adjusting the sequence length from $L$ to $P$.

First, the feature dimension is mapped using two sequential point-wise linear projections, similar to the CSI embedding layer but in reverse. 
Mathematically, this mapping step can be represented as:
\begin{equation}
    \mathbf{H}' = \text{Linear}_4 \left( \text{Linear}_3 \left( \mathbf{H}_{\text{LLM}} \right) \right),
\end{equation}
where $\mathbf{H}' \in \mathbb{R}^{B \times L \times N}$. $\text{Linear}_3$ maps features $E_g \rightarrow N'_{hs}$ and $\text{Linear}_4$ maps $N'_{hs} \rightarrow N$, with $N'_{hs}$ being an intermediate feature dimension.  

Second, the sequence length is transformed from $L$ to $P$. 
This is achieved by applying linear transformations across the time dimension independently for each of the $N$ features. 
To facilitate this, the tensor $\mathbf{H}'$ is first permuted to $(B, N, L)$. Then, two sequential linear layers project the time dimension:
\begin{equation}
    \hat{\mathbf{H}}_{\text{permuted}} = \text{Linear}_6 \left( \text{Linear}_5 \left( \mathbf{H}'_{\text{permuted}} \right) \right),
\end{equation}
where $\mathbf{H}'_{\text{permuted}} \in \mathbb{R}^{B \times N \times L}$ and $\hat{\mathbf{H}}_{\text{permuted}} \in \mathbb{R}^{B \times N \times P}$. $\text{Linear}_5$ maps the length $L \rightarrow N''_{hs}$ and $\text{Linear}_6$ maps $N''_{hs} \rightarrow P$, where $N''_{hs}$ is an intermediate dimension for the sequence length transformation. 
Finally, the resulting tensor is permuted back to the desired shape $\hat{\mathbf{H}}_{\text{norm}} \in \mathbb{R}^{B \times P \times N}$. 

Practically, all of these linear transformations are also efficiently implemented using $1\text{-D}$ convolutional layers.

\subsection{De-Normalization}
The final step involves transforming the normalized predicted sequence $\hat{\mathbf{H}}_{\text{norm}} \in \mathbb{R}^{B \times P \times N}$ back to the original scale of the input CSI data. 
This de-normalization process reverses the initial operation performed by the normalization layer. 
It utilizes the mean $\mu_{\mathbf{H}}$ and standard deviation $\sigma_{\mathbf{H}}$ computed and stored during the normalization step, as described in (\ref{eq:norm_layer}). 
The de-normalization is performed element-wise as follows:
\begin{equation} \label{eq:denorm_layer}
    \hat{\mathbf{H}}_{\text{pred}} = \hat{\mathbf{H}}_{\text{norm}} \cdot \sigma_{\mathbf{H}} + \mu_{\mathbf{H}} .
\end{equation}
% where the scalar values $\mu_{\mathbf{H}}$ and $\sigma_{\mathbf{H}}$ are broadcast across all dimensions (batch, prediction length, and feature) of $\hat{\mathbf{H}}_{\text{norm}}$. 
The resulting tensor $\hat{\mathbf{H}}_{\text{pred}} \in \mathbb{R}^{B \times P \times N}$ represents the final predicted future CSI sequence in its original scale.

\section{Numerical Results}
This section describes the environment setup, baseline methods, training details, evaluation results, and ablation studies \footnote{Our source code is available at \url{https://github.com/hele082/SCA-LLM}.}.

\subsection{Environment Setup}
To generate realistic time-varying channel data for training and evaluation, we use the widely adopted QuaDRiGa channel simulator \cite{jaeckel2014quadriga}, configured to generate datasets according to the 3GPP standards \cite{3gpp_tr38901_2018}. 
In particular, a MIMO-OFDM system is implemented in the TDD mode, and the BS and UE are equipped with $N_{BS} = 4$ and $N_{UE} = 4$ antennas, respectively. 
The aforementioned system uses an OFDM channel with the center frequency of $2.4~\text{GHz}$ and the bandwidth of $1.44~\text{MHz}$, containing $12$ subcarriers spaced by $\Delta f = 120~\text{KHz}$.  
The time interval between consecutive CSI samples, defined by the SRS periodicity, is $0.625~\text{ms}$.
For the channel prediction task, we set the historical sequence length to $L=24$ and the future prediction length to $P=6$.

In addition, the datasets are generated with a range of user mobility scenarios, specifically considering $13$ different user velocities uniformly distributed between $0~\text{km/h}$ and $60~\text{km/h}$. 
The models are trained and validated exclusively using data from the 3GPP UMa channel model under non-line-of-sight (NLOS). 
Specifically, the training dataset comprises $10400$ samples ($800$ per velocity), while the validation dataset contains $2600$ samples ($200$ per velocity). 
During the training, the SNR for each sample is randomly selected from the range $[0~\text{dB}, 20~\text{dB}]$ to enhance model robustness against varying noise levels.
To evaluate performance and generalization capabilities, we prepare distinct test datasets for two different scenarios:
\begin{itemize}
    \item \textbf{UMa NLOS Scenario:} This dataset serves to evaluate the model performance under conditions matching the training environment.
    \item \textbf{UMi NLOS Scenario:} This dataset assesses the model's ability to generalize to the 3GPP UMi channel model, which represents a different propagation environment not seen during the training.
\end{itemize}
The above two test datasets UMa and UMi consist of $13000$ samples each, containing $1000$ samples for every one of the $13$ velocities ($0~\text{km/h}$ to $60~\text{km/h}$). This comprehensive testing setup allows us to assess both in-distribution accuracy and cross-scenario robustness.

\subsection{Baseline Methods}
To comprehensively evaluate the performance of our proposed framework, we compare it against several relevant baseline methods. These include established sequence modeling architectures and the pioneering LLM4CP work \cite{DBLP:LLM4CP}, the first study to leverage LLMs for the channel prediction task. This method serves as a key LLM-based baseline in our evaluations. The specific baselines considered are:

\begin{itemize}
    \item \textbf{RNN}\cite{DBLP:RNN}: A RNN based method. RNN based methods are frequently used in time-series analysis and serve as a fundamental baseline for sequence processing tasks like channel prediction.
    \item \textbf{LSTM}\cite{DBLP:LSTM}: A LSTM based method. This method is specifically designed to capture long-range dependencies often present in time-varying channel data.
    \item \textbf{GRU}\cite{DBLP:GRU}: A GRU based method. It offers a variation on LSTMs, often providing comparable performance with greater computational efficiency.
    \item \textbf{Transformer}\cite{DBLP:Transformer}: A transformer based architecture adapted for the sequence-to-sequence channel prediction task. This represents a non-recurrent approach to sequence modeling.
    \item \textbf{LLM4CP}\cite{DBLP:LLM4CP}: The pioneering work that introduced the use of a fine-tuned pre-trained GPT-2 model for channel prediction in MISO-OFDM systems. We include LLM4CP as the key baseline representing the LLM-based approach in this area.
\end{itemize}

As to the network design of the proposed SCA-LLM framework, specific configurations are used in the experiments unless stated otherwise. The embedding size $E_{c}=64$ of the CSI embedding layer is used. The SCA module include $N_{ca}=4$ multi-spectral channel attention layers with an intermediate channel dimension $L_{c} = 128$. The DCT analysis dimensions $H_{dct}=W_{dct}=7$, and $n=32$ DCT bases are adopted. Moreover, the smallest GPT-2 variant with the first $N_{LLM} = 6$ decoder layers and embedding dimension $E_{g}=768$ is employed, and all convolutional layers share a hidden size of $128$ (i.e., $N_{hs}=N_{hs}'=N_{hs}''=128$).
The selection and configuration of RNN, LSTM, GRU, and Transformer mirrors the comparative analysis performed in the LLM4CP \cite{DBLP:LLM4CP}, facilitating a consistent evaluation context relative to prior work. For a fair comparison, the convolutional layers within the LLM4CP baseline is employed the same number of layers and hidden dimensions as our proposed SCA-LLM. The settings of GPT-2 model for LLM4CP is also the same.

\subsection{Training Details}
The proposed framework and the baselines are first trained using the same training dataset before their performance are assessed on the test datasets. 
During the training phase, each method utilizes the ground truth future CSI sequences, denoted as $\mathbf{H}_{\text{true}} \in \mathbb{R}^{B \times P \times N}$. 
The objective of the training is to reduce the difference between the model's output predictions, $\mathbf{H}_{\text{pred}}$, and these ground truth targets. This difference is measured by the NMSE, which serves as the loss function of:
\begin{equation} \label{metric:NMSE}
    Loss = \frac{|| \mathbf{H}_{\text{pred}} - \mathbf{H}_{\text{true}}||_{F}^{2}}{|| \mathbf{H}_{\text{true}}||_{F}^{2}}.
\end{equation}
The NMSE is also employed for calculating the validation loss during training. Ultimately, the version of each model exhibiting the lowest validation loss throughout the training process is preserved and utilized for the subsequent testing phase. 

For the proposed SCA-LLM framework, it is worth noting that our adaptation strategy involves a selective fine-tuning process for the pre-trained backbone, as illustrated in Fig.~\ref{fig:model_architecture}. Specifically, within the pre-trained GPT-2, only the parameters associated with the learnable positional embedding and LN layers are fine-tuned (made trainable), while the parameters of the MHSA and FFN layers remain fixed (frozen) to preserve the general sequence modeling capabilities learned during pre-training. The other parts within the proposed framework, namely the adapter and the output head, are fully trainable. Consequently, despite the other parts being fully trainable, the total number of trainable parameters in our framework remains significantly smaller than its overall parameter count, as the frozen core components of the pre-trained LLM backbone account for the vast majority of the total parameters.

% \subsection{Evaluation Metrics}
% We assess the performance of the proposed framework and the baselines using the following standard metrics:

% \begin{itemize}
%     \item \textbf{NMSE:} As defined in Eq.~\eqref{metric:NMSE}, NMSE is a widely adopted metric that directly quantifies the accuracy of the channel prediction by measuring the squared error relative to the true channel power. It serves as our primary performance indicator.
%     \item \textbf{Spectral Efficiency (SE):} SE measures the achievable data rate (e.g., in bits/s/Hz) given the predicted CSI. It provides a crucial communication-centric evaluation, reflecting how effectively the prediction translates to potential system throughput. Assuming downlink transmission where the BS employs conjugate beamforming (matched filtering), the SE is closely related to the alignment between the predicted and true channels, which is given by
%     \begin{equation} \label{metric:SE}
%         \text{SE} = \mathbb{E} \left[ \log_2 \left( 1 + \text{SNR} \cdot \frac{|\mathbf{h}_{pred}^H \mathbf{h}_{true}|^2}{||\mathbf{h}_{pred}||_F^2 ||\mathbf{h}_{true}||_F^2} \right) \right]
%     \end{equation}
%     where $\text{SNR}$ represents the average signal-to-noise ratio, and $\mathbb{E}[\cdot]$ denotes the expectation computed over all samples in the test set. This metric highlights how prediction accuracy impacts achievable rates.
% \end{itemize} 

\subsection{Evaluation Results}
In this part, we compare the performance of the proposed framework and the baselines using the NMSE metric in \eqref{metric:NMSE}, a widely adopted metric that directly quantifies the accuracy of the channel prediction by measuring the normalized mean squared error relative to the true channel power. 
A smaller NMSE indicates a more accurate channel prediction, i.e., lower is better. In the comparisons, our proposed SCA-LLM framework is denoted as \textbf{Ours}.

\subsubsection{NMSE Performance Comparison under Varying SNR}
\begin{figure*}[!htbp] % Use figure* to span two columns
    \centering % Center the subfigures horizontally

    % --- Subfigure (a) for UMa 0km/h ---
    \subfloat[UMa NLOS scenario (trained from scratch)\label{fig:NMSE_vs_SNR_0kmh_UMa}]{%
        \includegraphics[width=0.48\linewidth]{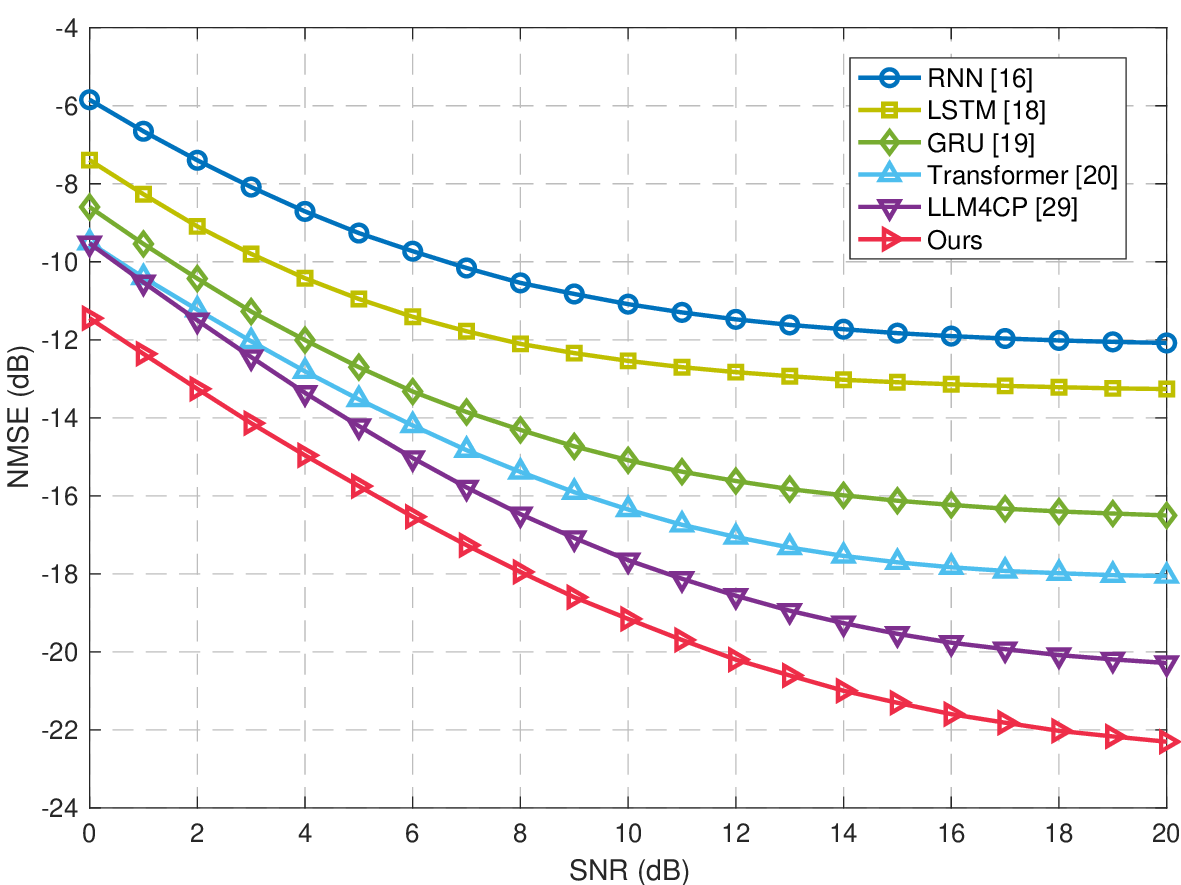}%
    }
    \hfill % Space between subfigures
    % --- Subfigure (b) for UMi 0km/h ---
    \subfloat[UMi NLOS scenario (zero-shot evaluation; trained on UMa)\label{fig:NMSE_vs_SNR_0kmh_UMi}]{%
        \includegraphics[width=0.48\linewidth]{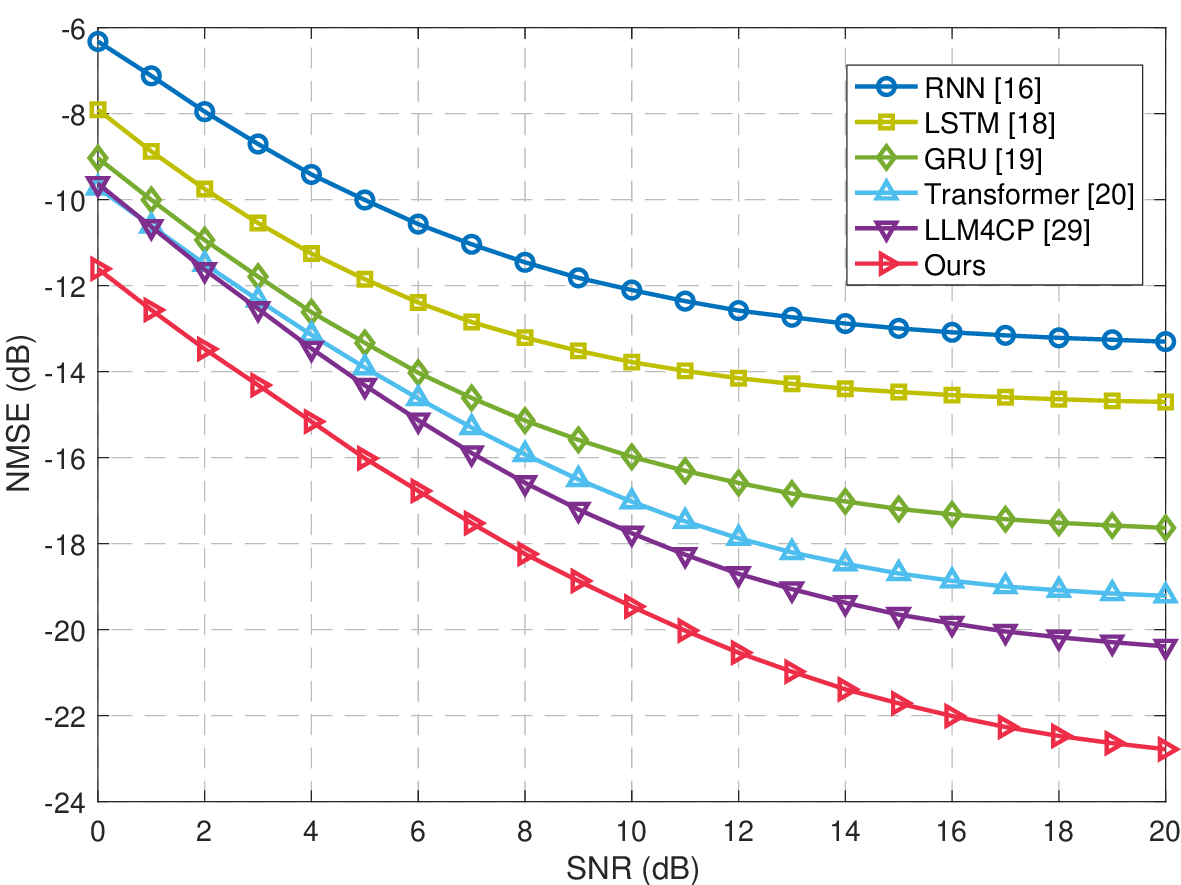}%
    }

    % --- Main Caption and Label for Figure ---
    \caption{NMSE performance versus SNR between UMa and UMi NLOS scenarios with $\text{the UE velocity} = 0~\text{km/h}$.}
    \label{fig:NMSE_vs_SNR_0kmh} % Label for the WHOLE Figure
\end{figure*}

\begin{figure*}[!htbp] % Use figure* to span two columns
    \centering % Center the subfigures horizontally

    % --- Subfigure (a) for UMa 50km/h ---
    \subfloat[UMa NLOS scenario (trained from scratch)\label{fig:NMSE_vs_SNR_50kmh_UMa}]{%
        \includegraphics[width=0.48\linewidth]{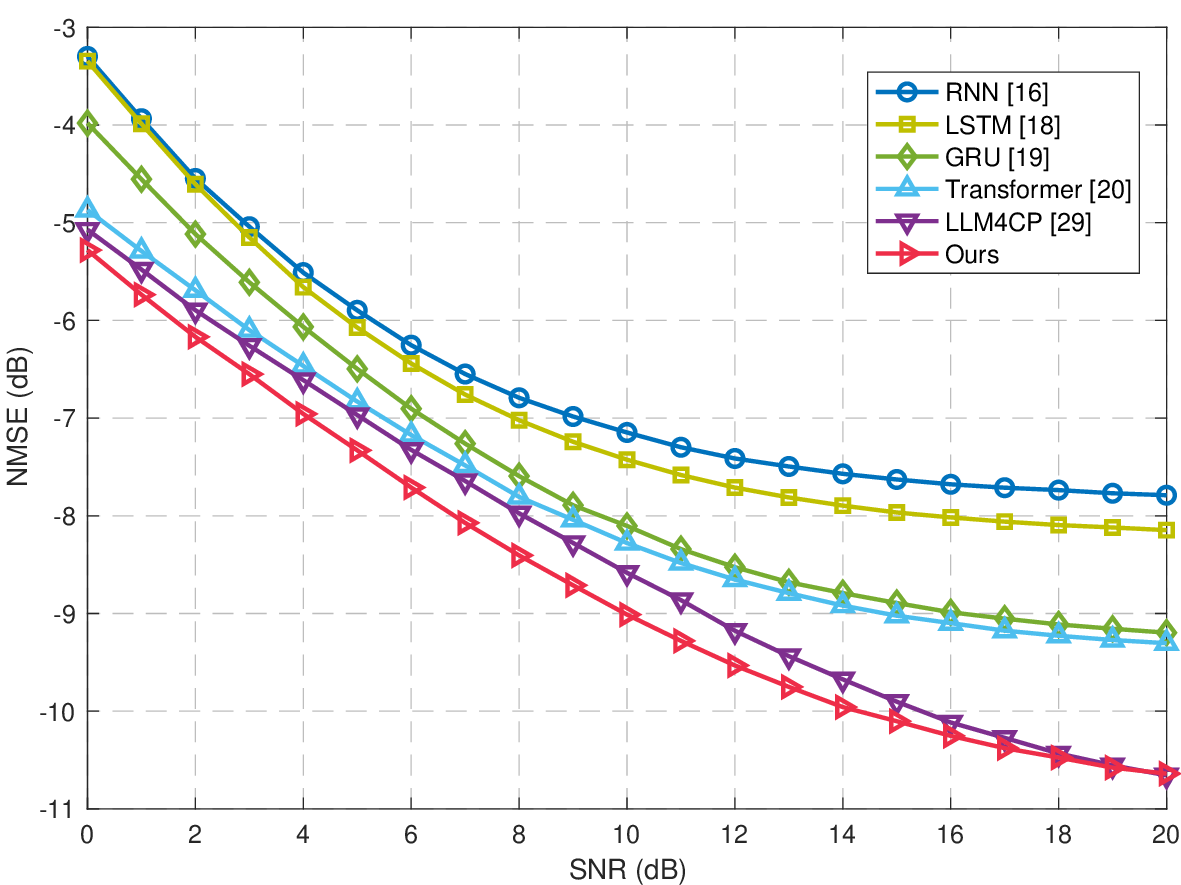}%
    }
    \hfill % Space between subfigures
    % --- Subfigure (b) for UMi 50km/h ---
    \subfloat[UMi NLOS scenario (zero-shot evaluation; trained on UMa)\label{fig:NMSE_vs_SNR_50kmh_UMi}]{%
        \includegraphics[width=0.48\linewidth]{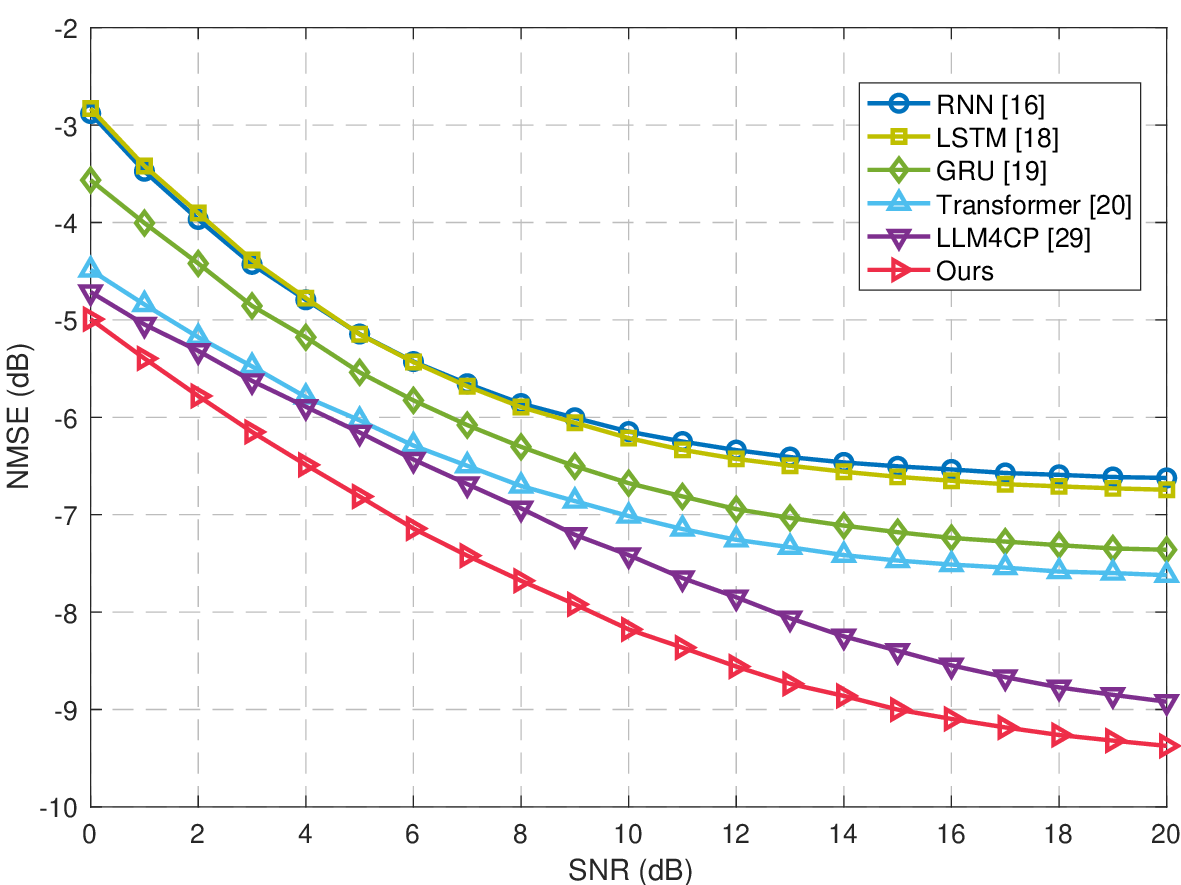}%
    }

    % --- Main Caption and Label for Figure ---
    \caption{NMSE performance versus SNR between UMa and UMi NLOS scenarios with $\text{the UE velocity} = 50~\text{km/h}$.}
    \label{fig:NMSE_vs_SNR_50kmh} % Label for the WHOLE Figure
\end{figure*}

We evaluate the NMSE performance of our proposed framework against several baselines as the SNR varies. 
The results for UE velocities of $0~\text{km/h}$ and $50~\text{km/h}$ are presented in Fig.~\ref{fig:NMSE_vs_SNR_0kmh} and Fig.~\ref{fig:NMSE_vs_SNR_50kmh}, respectively, covering both UMa and UMi NLOS scenarios. 
As depicted in these figures, a general trend is observed across all methods and scenarios: the NMSE decreases significantly (indicating improved prediction accuracy) as the SNR increases from $0~\text{dB}$ to $20~\text{dB}$. 
This aligns with the expectation that higher SNR levels facilitate more accurate channel prediction. 

Focusing first on the static scenario ($0~\text{km/h}$), which is crucial for applications like fixed wireless access or low-mobility IoT devices, Fig.~\ref{fig:NMSE_vs_SNR_0kmh} details the performance comparison. 
In the UMa NLOS scenario (Fig.~\ref{fig:NMSE_vs_SNR_0kmh_UMa}), corresponding to the training distribution, our proposed method significantly outperforms all baselines. 
For instance, at an SNR of $20~\text{dB}$, our method demonstrates a performance advantage of approximately $-2.0~\text{dB}$ over the next best baseline, LLM4CP (achieving NMSEs of roughly $-22.3~\text{dB}$ versus $-20.3~\text{dB}$, respectively), and is substantially better than the traditional methods (RNN, LSTM, GRU, Transformer). 

When evaluating the generalization capability in the unseen UMi NLOS scenario (Fig.~\ref{fig:NMSE_vs_SNR_0kmh_UMi}), our method maintains its clear performance leadership. 
Interestingly, while all methods tend to exhibit degradation in NMSE performance when generalizing from the trained UMa environment to the unseen UMi environment, our method still consistently achieves the lowest NMSE across the tested SNR range. 
Furthermore, the performance gap between our method and LLM4CP widens in the zero-shot UMi setting.
In particular, at $20~\text{dB}$ SNR, this gap increases from approximately $-2.0~\text{dB}$ in UMa to $-2.4~\text{dB}$ in UMi, where the respective NMSEs are $-22.7~\text{dB}$ (our method) and $-20.3~\text{dB}$ (LLM4CP).
This further highlights the superior robustness and generalization of our proposed framework.

Turning to the high-mobility scenario represented by a UE velocity of $50~\text{km/h}$, Fig.~\ref{fig:NMSE_vs_SNR_50kmh} presents the corresponding NMSE versus SNR results. 
As expected, the faster channel variations at this higher speed lead to increased prediction difficulty, generally resulting in a noticeable degradation of NMSE performance across all methods compared to the static case. 
At the UE velocity of $50~\text{km/h}$, transitioning from the trained UMa scenario to the unseen UMi scenario leads to a discernible performance degradation (increased NMSE) across all methods. 
This contrasts with the static case, where generalizing to the UMi scenario actually resulted in slight performance improvements for all methods.
However, the fundamental performance patterns observed at $0~\text{km/h}$ are still persist. 
Our method continues to deliver the best performance in both the UMa NLOS (Fig.~\ref{fig:NMSE_vs_SNR_50kmh_UMa}) and the UMi NLOS (Fig.~\ref{fig:NMSE_vs_SNR_50kmh_UMi}) scenarios. 
For instance, in the UMa scenario at $10~\text{dB}$ SNR, our method demonstrates a clear advantage.
It outperforms LLM4CP by approximately $-0.5~\text{dB}$ and also surpasses other baselines. 
Furthermore, this performance advantage widens in the zero-shot UMi setting. Here, at $10~\text{dB}$ SNR, our method leads LLM4CP by about $-0.8~\text{dB}$.
This larger performance lead over LLM4CP in the UMi scenario shows the robust generalization of our method, even under challenging high-mobility conditions.

\subsubsection{NMSE Performance Comparison under Varying UE Velocity}
\begin{figure*}[!htbp] % Use figure* to span two columns
    \centering % Center the subfigures horizontally

    % --- Subfigure (a) for UMa ---
    \subfloat[UMa NLOS scenario (trained from scratch)\label{fig:NMSE_vs_Velocity_10dB_UMa}]{%
        \includegraphics[width=0.48\linewidth]{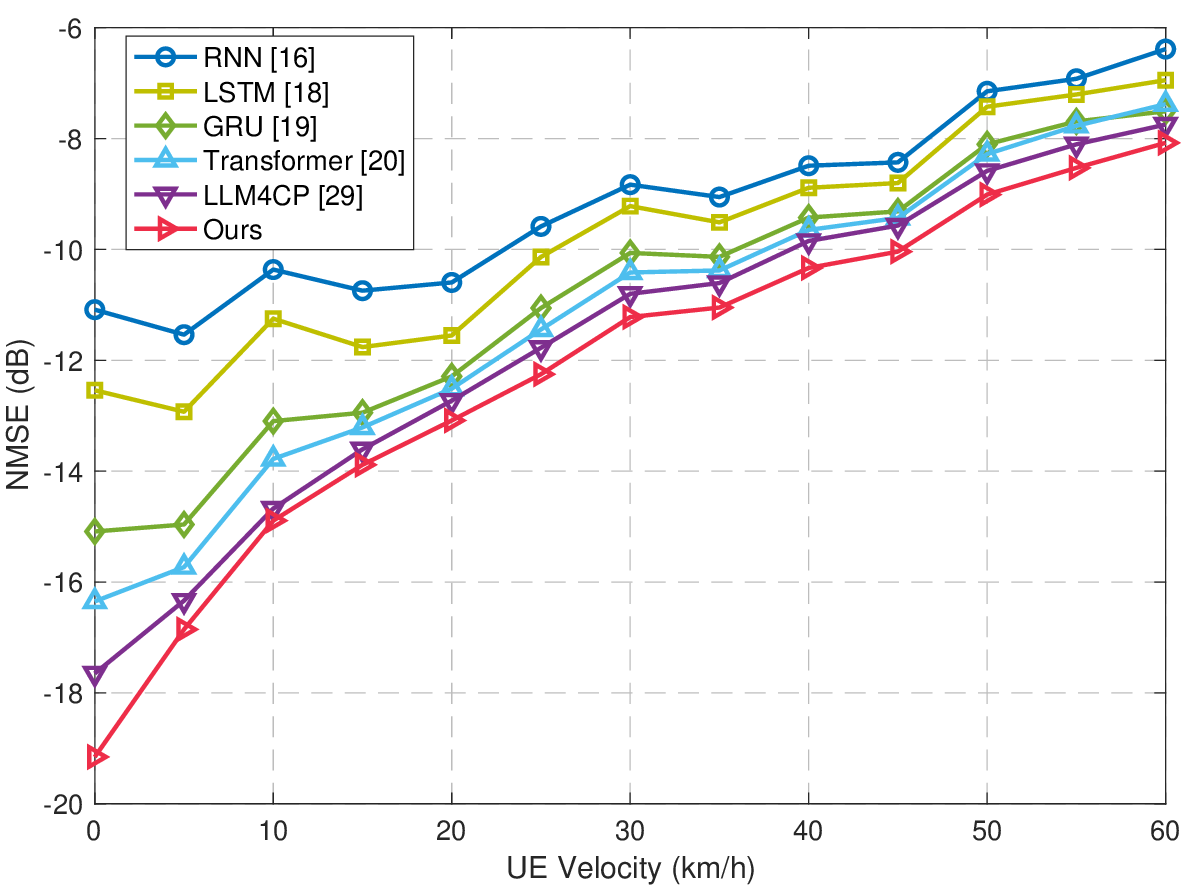}%
    }
    \hfill % Space between subfigures
    % --- Subfigure (b) for UMi ---
    \subfloat[UMi NLOS scenario (zero-shot evaluation; trained on UMa)\label{fig:NMSE_vs_Velocity_10dB_UMi}]{%
        \includegraphics[width=0.48\linewidth]{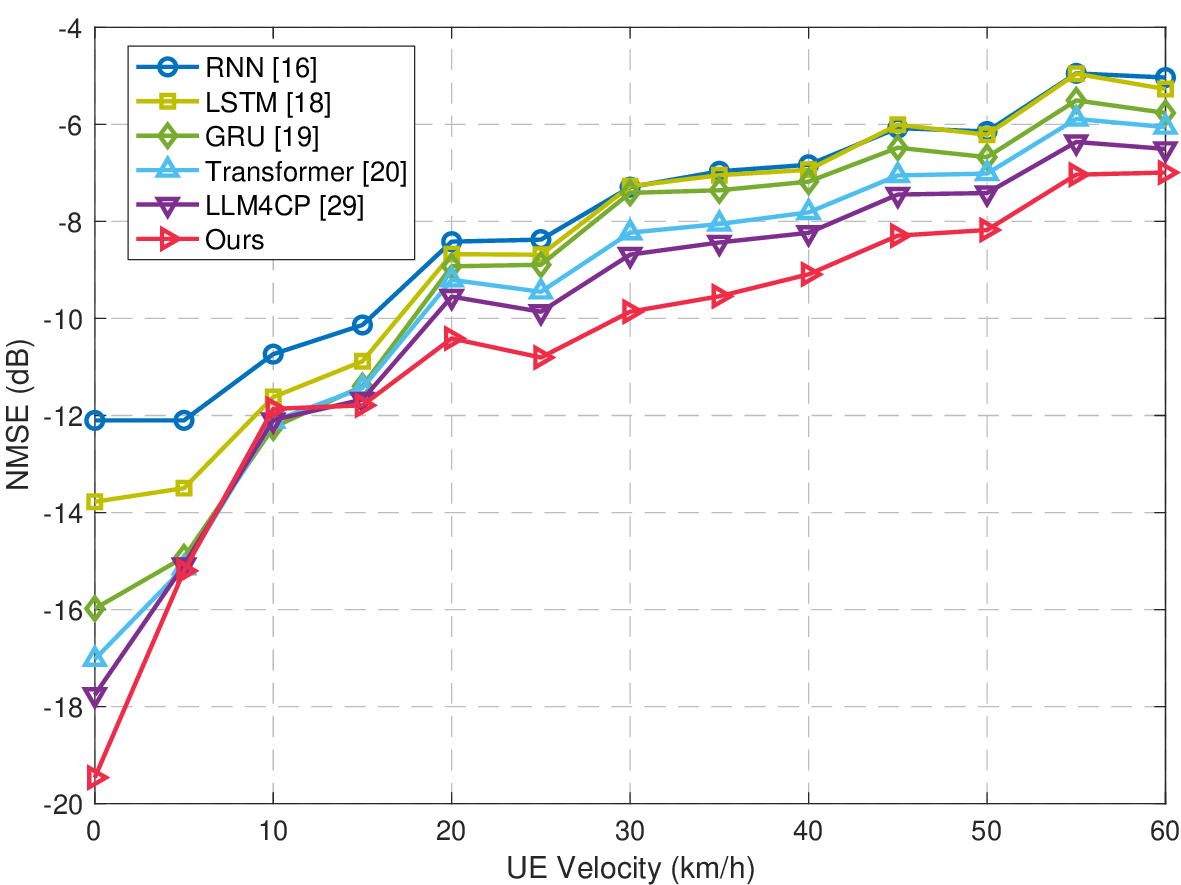}%
    }

    % --- Main Caption and Label for Figure ---
    % Caption updated to match the template style
    \caption{NMSE performance versus UE velocity between UMa and UMi NLOS scenarios with $\text{SNR}=10~\text{dB}$.}
    \label{fig:NMSE_vs_Velocity_10dB} % Label for the WHOLE Figure
\end{figure*}
To further investigate the impact of user mobility on channel prediction accuracy, we analyze the NMSE performance across the full range of simulated UE velocities ($0~\text{km/h}$ to $60~\text{km/h}$) at a fixed $\text{SNR} = 10~\text{dB}$. Fig.~\ref{fig:NMSE_vs_Velocity_10dB} presents these results for both UMa and UMi NLOS scenarios.
A clear trend observed in both scenarios is that the performance generally degrades (NMSE increases) for all methods as the UE velocity rises. This is expected, as higher velocities induce faster channel variations, making accurate prediction more challenging. The rate of performance degradation tends to be steeper at lower velocities and somewhat flattens as the velocity approaches $60~\text{km/h}$.

In the UMa NLOS scenario (Fig.~\ref{fig:NMSE_vs_Velocity_10dB_UMa}), which aligns with the training conditions, our proposed method consistently demonstrates superior performance, achieving the lowest NMSE across the entire velocity spectrum. 
Specifically, at $0~\text{km/h}$, our method achieves an NMSE of approximately $-19.2~\text{dB}$, outperforming the LLM4CP baseline (around $-17.6~\text{dB}$) by about $-1.6~\text{dB}$. 
As velocity increases, while the absolute performance of all methods degrades, our framework still maintains its advantage. For instance, at the UE velocities of $30~\text{km/h}$ and $60~\text{km/h}$, our method leads LLM4CP by approximately $-0.4~\text{dB}$ and $-0.3~\text{dB}$, respectively.

Turning to the generalization performance in the unseen UMi NLOS scenario (Fig.~\ref{fig:NMSE_vs_Velocity_10dB_UMi}), our proposed framework again maintains its clear lead over all baselines across all tested velocities. Similar to the observations in the SNR analysis, the performance gap between our method and LLM4CP is even more pronounced in this zero-shot scenario compared to the UMa case. At $0~\text{km/h}$, the advantage over LLM4CP increases to about $-1.7~\text{dB}$. This enhanced lead persists at higher velocities, with gaps of approximately $-1.2~\text{dB}$ at $30~\text{km/h}$ and $-0.5~\text{dB}$ at $60~\text{km/h}$. These results again underscore the robustness and superior generalization capability of our proposed method across varying mobility conditions.

\subsubsection{NMSE Performance per Prediction Step Comparison}

\begin{figure*}[!htbp] % Use figure* to span two columns
    \centering % Center the subfigures horizontally

    % --- Subfigure (a) for UMa ---
    \subfloat[UMa NLOS scenario  (trained from scratch)\label{fig:NMSE_vs_Prediction_Length_50Kmh_10dB_UMa}]{%
        \includegraphics[width=0.48\linewidth]{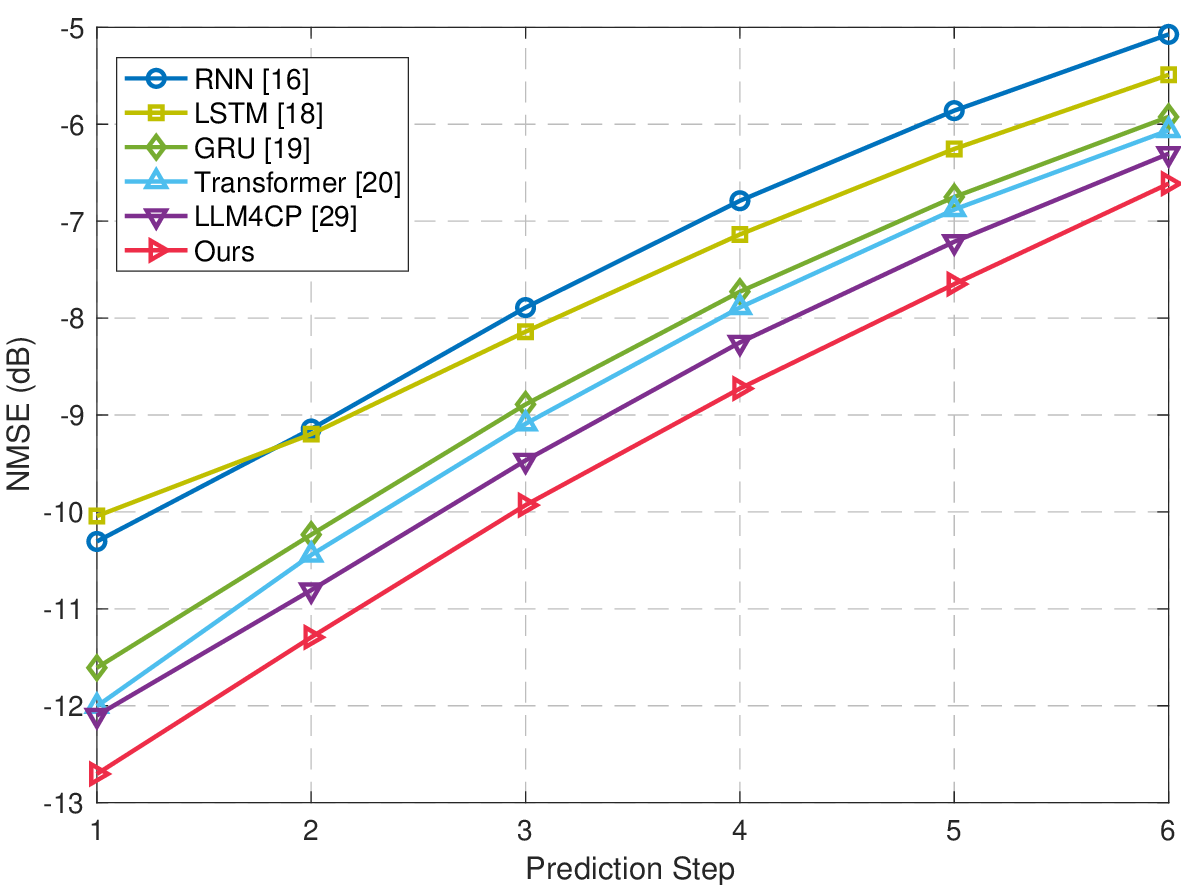}%
    }
    \hfill % Space between subfigures
    % --- Subfigure (b) for UMi ---
    \subfloat[UMi NLOS scenario (zero-shot evaluation; trained on UMa)\label{fig:NMSE_vs_Prediction_Length_50Kmh_10dB_UMi}]{%
        \includegraphics[width=0.48\linewidth]{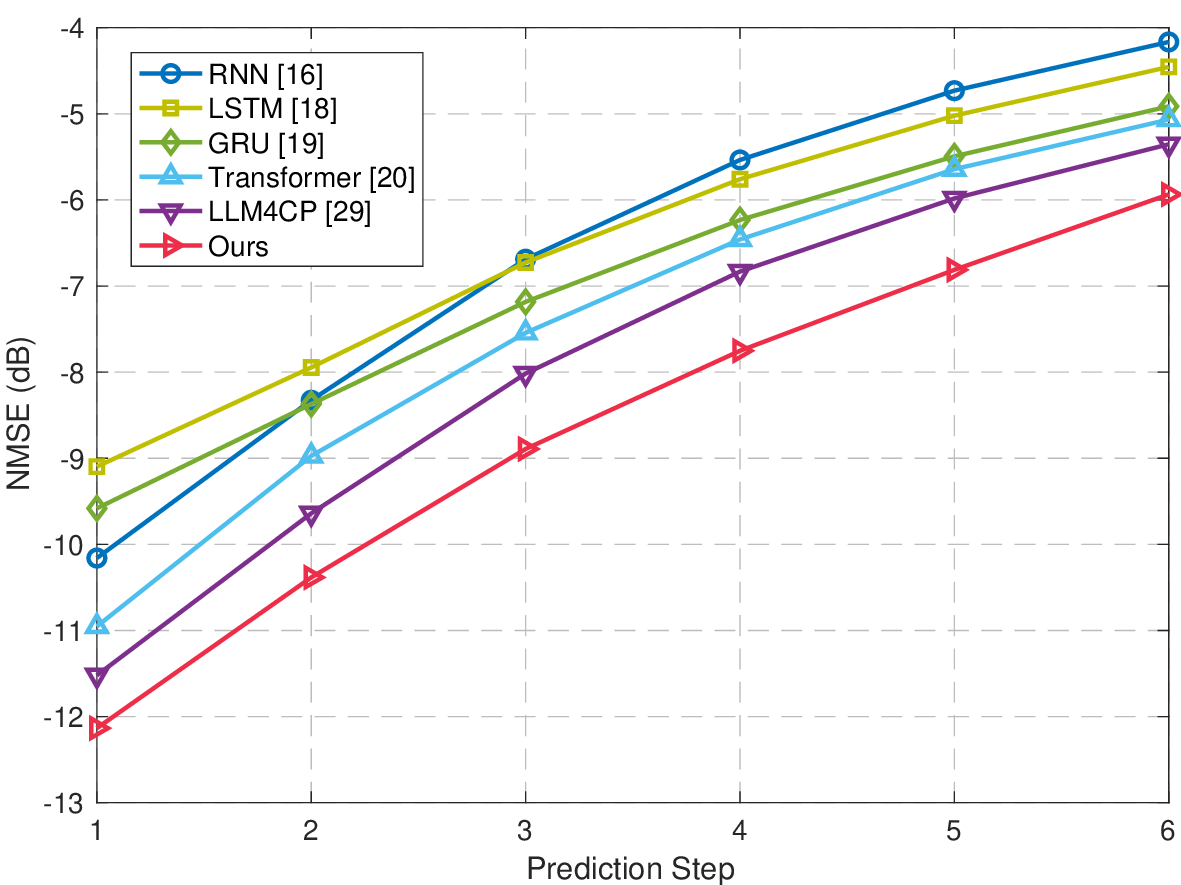}%
    }

    % --- Main Caption and Label for Figure ---
    % Caption updated to match the template style
    \caption{NMSE performance versus prediction step $p$ between UMa and UMi NLOS scenarios with $\text{UE velocity}=50~\text{km/h}$ and $\text{SNR}=10~\text{dB}$. The total prediction horizon is $P=6$.}
    \label{fig:NMSE_vs_Prediction_Length_50Kmh_10dB} % Label for the WHOLE Figure
\end{figure*}
In this part, we delve deeper into the prediction performance by examining the NMSE specifically at each prediction step within the total prediction horizon $P=6$. 
Fig.~\ref{fig:NMSE_vs_Prediction_Length_50Kmh_10dB} presents this step-wise NMSE evaluation under challenging conditions ($\text{UE velocity}=50~\text{km/h}$, $\text{SNR}=10~\text{dB}$) for both UMa and UMi scenarios. 
In this figure, the prediction step ranges from the first step ($p=1$) to the last step ($p=6$).
As expected, NMSE performance degrades with the prediction step for all methods in both scenarios, as predicting further into the future inherently involves a larger uncertainty and error accumulation.

In the UMa NLOS scenario (Fig.~\ref{fig:NMSE_vs_Prediction_Length_50Kmh_10dB_UMa}), our proposed method consistently achieves the lowest NMSE at every prediction step. 
For the first prediction step with $p=1$, our method yields an NMSE of approximately $-12.7~\text{dB}$, surpassing LLM4CP ($\sim -12.1~\text{dB}$) by about $-0.6~\text{dB}$ and significantly lower than the other baselines. 
This performance advantage is sustained across the prediction horizon, with the gap relative to LLM4CP ranging between $-0.6~\text{dB}$ and $-0.3~\text{dB}$ for steps $p=1$ to $p=6$. 

Analyzing the generalization performance in the unseen UMi NLOS scenario (Fig.~\ref{fig:NMSE_vs_Prediction_Length_50Kmh_10dB_UMi}), our method maintains its superiority at each prediction step. 
While all methods exhibit slightly higher NMSE compared to the UMa scenario due to the zero-shot evaluation context, our method's lead over the LLM4CP baseline is consistently maintained and generally more pronounced than that in the UMa case. 
In particular, for the first prediction step ($p=1$), the NMSE advantage over LLM4CP is about $-0.6~\text{dB}$. 
This performance gap widens for intermediate prediction steps, reaching a maximum advantage of approximately $-0.9~\text{dB}$ at steps $p=3$ and $p=4$, before settling back to a $-0.6~\text{dB}$ advantage when $p=6$. 
This sustained lead across the prediction horizon in both scenarios indicates the robustness and effectiveness of our proposed method for multi-step channel prediction.

\subsection{Ablation Studies}
\begin{figure}[htbp]
    \centering
    \includegraphics[width=0.5\linewidth]{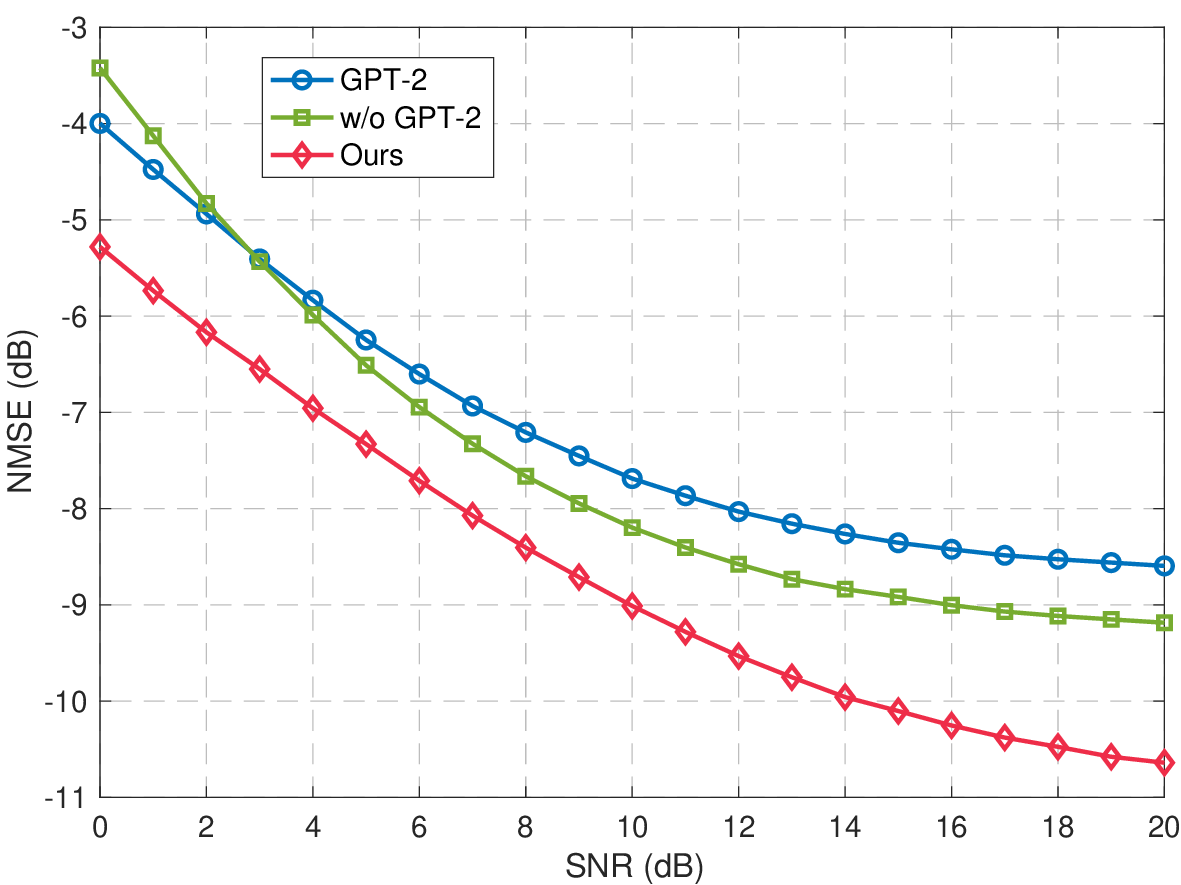}
    \caption{NMSE performance versus SNR in the UMa NLOS scenario with $\text{UE velocity}=50~\text{km/h}$.}
    \label{fig:Ablation_UMa_NMSE_vs_SNR_50Kmh}
\end{figure}

\begin{figure}[htbp]
    \centering
    \includegraphics[width=0.5\linewidth]{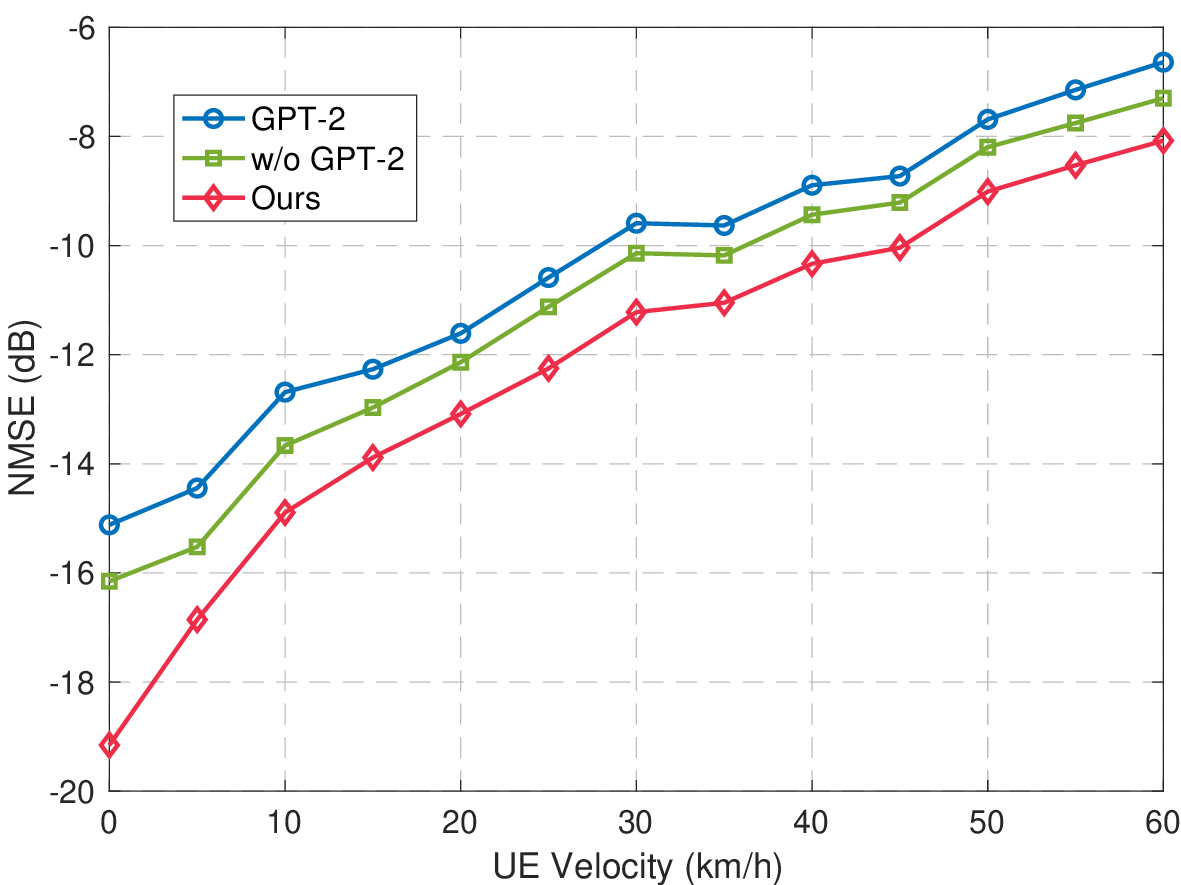}
    \caption{NMSE performance versus UE velocity in the UMa NLOS scenario with $\text{SNR}=10~\text{dB}$.}
    \label{fig:Ablation_UMa_NMSE_vs_Velocity_10dB}
\end{figure}

\begin{figure}[htbp]
    \centering
    \includegraphics[width=0.5\linewidth]{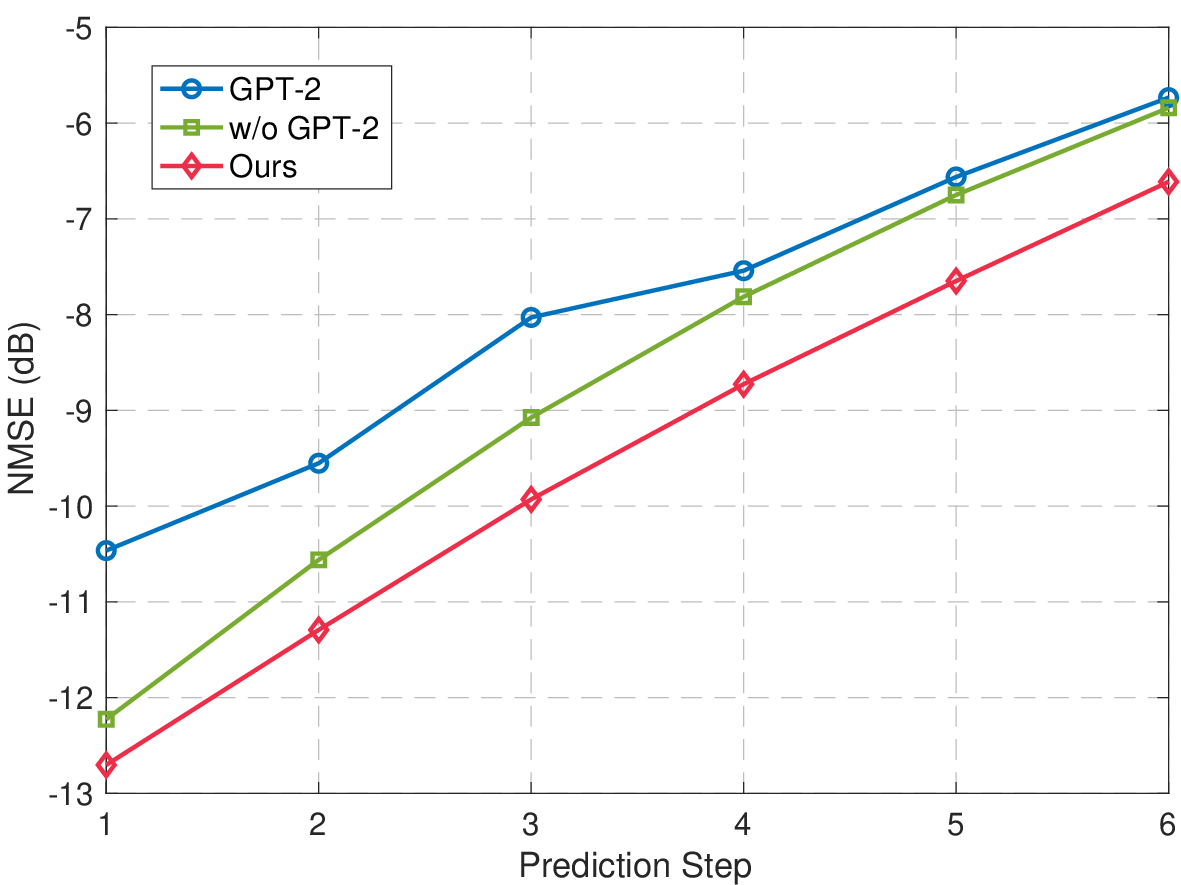}
    \caption{NMSE performance versus prediction step $p$ in the UMa NLOS scenario with $\text{UE velocity}=50~\text{km/h}$ and $\text{SNR}=10~\text{dB}$. The total prediction horizon is $P=6$.}
    \label{fig:Ablation_UMa_NMSE_vs_Prediction_Length_50Kmh_10dB}
\end{figure}

To validate the effectiveness and necessity of the key components within our proposed framework, we conduct several ablation studies comparing our full proposed method against two variants:
\begin{itemize}
    \item \textbf{w/o GPT-2:} This variant removes the entire LLM component (pre-trained LLM backbone and its embedding), relying solely on the proposed adapter (i.e., CSI embedding and SCA module) and the output head. Its primary purpose is to establish the performance baseline achieved without the LLM component, thereby directly demonstrating the indispensable role of the LLM backbone.
    \item \textbf{GPT-2:} This variant removes the \textit{adapter} and instead processes the normalized CSI data directly by GPT-2 with full fine-tuning (FFT). The motivation is to create a strong baseline by evaluating a fully fine-tuned GPT-2 adapting all parameters directly to the task, without our specialized adapter. Comparing our method against this challenging baseline underscores the critical necessity and contribution of our specialized adapter.
\end{itemize}
The performance comparisons under the UMa NLOS scenario are presented in Fig.~\ref{fig:Ablation_UMa_NMSE_vs_SNR_50Kmh} (vs SNR), Fig.~\ref{fig:Ablation_UMa_NMSE_vs_Velocity_10dB} (vs UE velocity), and Fig.~\ref{fig:Ablation_UMa_NMSE_vs_Prediction_Length_50Kmh_10dB} (vs prediction step). In these figures, the full version of our proposed method is denoted as \textbf{Ours}.

First, the results clearly establish the importance of the LLM backbone. Comparing our proposed method with the ``w/o GPT-2'' variant reveals a significant performance degradation across all conditions (Figs.~\ref{fig:Ablation_UMa_NMSE_vs_SNR_50Kmh}-\ref{fig:Ablation_UMa_NMSE_vs_Prediction_Length_50Kmh_10dB}) when the LLM is removed. 
This gap is substantial and notably widens for predictions further into the future (Fig.~\ref{fig:Ablation_UMa_NMSE_vs_Prediction_Length_50Kmh_10dB}).
This demonstrates the crucial role of the powerful sequence modeling capabilities of the pre-trained LLM in achieving high channel prediction performance.
However, while LLM is indispensable, it is not sufficient for optimal performance in the channel prediction task. 
This is evident when comparing our proposed method against the ``GPT-2'' variant. 
Our method consistently and significantly outperforms ``GPT-2'' across all scenarios, despite the latter employing FFT across all its parameters. 
This substantial performance advantage powerfully underscores the critical importance and effectiveness of our proposed adapter. 
It indicates the adapter can effectively preprocess and transform the CSI data into a representation that the LLM can readily comprehend and model, thereby unlocking its predictive potential within the proposed framework. 
In addition, it also indicates that simply applying an LLM directly to the channel prediction task, even with full fine-tuning, is insufficient for achieving the promising performance.

This insight is further emphasized by the comparison between the two ablated versions. 
Strikingly, the ``w/o GPT-2'' variant consistently outperforms the ``GPT-2'' variant across nearly all conditions (Figs.~\ref{fig:Ablation_UMa_NMSE_vs_SNR_50Kmh}-\ref{fig:Ablation_UMa_NMSE_vs_Prediction_Length_50Kmh_10dB}, except potentially at extremely low SNR). 
This compelling result strongly suggests that naive application of even a powerful, fully fine-tuned LLM yields mediocre performance, as the model struggles without domain-specific guidance. 
In contrast, it highlights the remarkable effectiveness of our proposed adapter in extracting useful information from CSI data, proving its value even in isolation.

Most importantly, our full method significantly outperforms the separated components across all tested SNRs, UE velocities, and prediction steps. 
This superiority highlights the effectiveness and superiority of the overall SCA-LLM framework. 
Within this framework, the LLM is adapted to achieve the SOTA performance compared to the baselines.
This results compellingly validates that the proposed adapter can bridge the gap between the CSI data and the LLM more effectively.

\section{Conclusions}
This paper proposed the SCA-LLM framework, which is a large-scale wireless world model for accurate and robust multi-step CSI prediction in MIMO-OFDM systems.
The framework effectively adapted pre-trained LLMs to CSI data by mitigating domain mismatch through a novel spectral-attentive adapter.
In this adapter, the SCA module employed multi-spectral channel attention with the 2-D DCT to capture a broader range of spectral components from CSI features compared to methods relying on global pooling operations.
This, in turn, provided a more suitable input for the LLM and significantly improved its adaptation and prediction capabilities.
Extensive simulations demonstrated that SCA-LLM achieved superior prediction performance, consistently outperformed existing methods, and yielded up to a notable NMSE advantage of $-2.4~\text{dB}$ over the prior LLM-based approach.
Ablation studies further validated the critical contributions of both the spectral-attentive adapter and the LLM backbone.
These findings underscored that the SCA-LLM framework strengthened the ``adapter + LLM'' paradigm, enabled better exploitation of LLMs for channel prediction, and suggested potential for broader LLM applications in wireless communications.
% This work addressed the problem of accurate, robust channel prediction in MIMO-OFDM systems, focusing on adapting pre-trained LLMs effectively. While LLMs offer powerful sequence modeling, their direct application is hindered by domain mismatch. To overcome this, the SCA-LLM framework was developed, integrating an LLM with a carefully designed spectral-attentive adapter. The key component of the adapter is the SCA module, which leverages multi-spectral channel attention based on the 2-D DCT. This technique preserves a richer set of spectral components from CSI features compared to conventional global average pooling, providing a more suitable input representation for the LLM and effectively bridging the domain gap. 
% Extensive numerical results demonstrated SCA-LLM's superiority, and ablation studies further validated the design by confirming the indispensable roles of both the adapter and the LLM. 
% % It achieved the SOTA prediction performance, outperforming established DL models and the prior LLM-based approach (LLM4CP) across different scenarios, various SNRs and user velocities. 
% % highlighting the spectral-attentive adapter's significant contribution to mitigating domain mismatch and enabling high performance. 
% All of these findings emphasize the effectiveness of the ``adapter + LLM'' paradigm in unlocking the full potential of LLMs for channel prediction, and pave the way for future research on the application of LLMs in wireless communications.

\begin{appendices}
\numberwithin{equation}{section}
\section{Multi-Spectral Channel Attention Layer} \label{layer: msca}
The multi-spectral channel attention layer implements a variant of the channel attention mechanism, a technique widely used in deep learning to adaptively recalibrate feature responses \cite{DBLP:SENet, DBLP:FcaNet, DBLP:Attention_Mechanism}. 
It is crucial to note the distinction in terminology: in the context of this attention mechanism, the term \textit{channel} refers to the \textit{channel} dimension within the data tensor (e.g., the dimension $L_c$ in $\mathbf{H}_{c}^{(i)} \in \mathbb{R}^{B \times L_c \times N \times E_c}$), and should be distinguished from the physical wireless communication channel. 
Conversely, the dimensions $(N, E_c)$ correspond to the spatial dimensions or the feature map upon which the subsequent analysis operates.

To incorporate richer information into the channel attention mechanism, this layer leverages the 2-D DCT \cite{DBLP:JPEG_DCT} for feature analysis and modeling. 
For the purpose of DCT analysis, each feature map is considered to have dimensions $H' \times W'$. These dimensions might correspond directly to the input spatial dimensions $(N, E_c)$, or they might be target dimensions $(H_{dct}, W_{dct})$ achieved through an optional pooling step described later. 
The 2-D DCT decomposes such a feature map into a sum of cosine functions oscillating at different frequencies. These basis functions form a complete orthogonal basis set, denoted as $\mathcal{B}_{\text{DCT}} = \{ \mathbf{F}_{u,v} | 0 \le u < H', 0 \le v < W' \}$. 
The standard basis function $\mathbf{F}_{u,v} \in \mathcal{B}_{\text{DCT}}$ corresponding to the frequency indices $(u,v)$ for this $H' \times W'$ map is defined as:
\begin{equation} \label{eq:dct_basis}
    \mathbf{F}_{u,v}[h,w] = \cos\left(\frac{\pi u}{H'}(h + \frac{1}{2})\right) \cos\left(\frac{\pi v}{W'}(w + \frac{1}{2})\right),
\end{equation}
where $ (h,w) $ are the spatial indices $ (0 \le h < H', 0 \le w < W') $ and $ (u,v) $ are the frequency indices identifying each basis function within the set $\mathcal{B}_{\text{DCT}}$.

The projection of a generic $H' \times W'$ feature map, let's denote it as $\mathbf{Q} \in \mathbb{R}^{H' \times W'}$ (representing one such map from the input tensor, potentially after pooling), onto a specific frequency component $(u,v)$ is achieved by calculating the corresponding DCT coefficient using the basis function $\mathbf{F}_{u,v}$. This projection is calculated as:
\begin{equation} \label{eq:dct_projection}
    \text{2DDCT}^{u,v}(\mathbf{Q}) = \sum_{h=0}^{H'-1} \sum_{w=0}^{W'-1} \mathbf{Q}[h,w] \cdot \mathbf{F}_{u,v}[h,w],
\end{equation}
where $\mathbf{Q}[h,w]$ denotes the element of $\mathbf{Q}$ at spatial index $(h,w)$.
Given the input tensor for the $i$-th layer, $ \mathbf{H}_{c}^{(i)} \in \mathbb{R}^{B \times L_c \times N \times E_c} $, where $ L_c $ is the previously defined channel dimension, the layer performs the following steps:

\subsection{Adaptive Pooling (Optional)}
If the input spatial dimensions $ (N, E_c) $ differ from the target DCT analysis dimensions $ (H_{dct}, W_{dct}) $, adaptive average pooling is applied independently to each $L_c$ channel map to obtain $ \mathbf{H}_{c,\text{pooled}}^{(i)} \in \mathbb{R}^{B \times L_c \times H_{dct} \times W_{dct}} $. After that, the DCT analysis dimensions would be $H'=H_{dct}, W'=W_{dct}$.

\subsection{Channel Splitting and DCT Projection}
The $ L_c $ input channels are divided into $ n $ groups, with $ L'_c = L_c / n $ channels per group. A specific 2-D DCT frequency index $ (u_k, v_k) $, corresponding to a basis function $\mathbf{F}_{u_k, v_k} \in \mathcal{B}_{\text{DCT}}$, is assigned to each group $ k \in \{0, \dots, n-1\} $. These indices identify a pre-selected subset of $n$ DCT basis functions, often chosen based on empirical findings for optimal performance as suggested in \cite{DBLP:FcaNet}.
It is worth mentioning that using these multiple pre-defined DCT basis functions adds negligible computational cost and no extra trainable parameters compared to using only the lowest frequency component via GAP \cite{DBLP:FcaNet}.
    
Let $ \mathbf{H}_{c,k}^{(i)} \in \mathbb{R}^{B \times L'_c \times H' \times W'} $ represent the feature maps for the $k$-th group (after optional pooling). The spectral features for this group are computed by applying the assigned 2-D DCT projection (Eq.~\eqref{eq:dct_projection}) using the basis function $\mathbf{F}_{u_k, v_k}$ to the $L'_c$ feature maps within the group $k$ independently:
    \begin{equation} \label{eq:fcanet_part_compression_redef}
         \mathbf{Freq}^k = \text{2DDCT}^{u_k, v_k}(\mathbf{H}_{c,k}^{(i)}) 
    \end{equation}
 This calculation results in a tensor $\mathbf{Freq}^k \in \mathbb{R}^{B \times L'_c}$, where each element represents the magnitude of the $(u_k, v_k)$ frequency component for a specific channel map within group $k$.

\subsection{Multi-Spectral Vector Concatenation} 
The frequency components calculated for all $n$ distinct groups are concatenated along the channel dimension, denoted as $Concat(\cdot)$. This aggregation forms a comprehensive multi-spectral descriptor $\mathbf{Y}^{(i)}_{\text{dct}}$ for each sample in the batch:
\begin{equation} \label{eq:fcanet_full_compression_redef}
\begin{split}
     \mathbf{Y}^{(i)}_{\text{dct}} = Concat([\mathbf{Freq}^0, &\mathbf{Freq}^1, \\ &\dots, \mathbf{Freq}^{n-1}]) \in \mathbb{R}^{B \times L_c}.
\end{split}
\end{equation}
These frequency components can help capture certain wireless channel characteristics.
% For example, low-frequency components may reflect the dominant spatial correlation structure across the antenna array arising from large clusters of scatterers or the line-of-sight path, while high-frequency components might capture fine-grained spatial variations and rapid fluctuations caused by local scattering, multipath interference, or subtle Doppler shifts from mobility.
        
\subsection{Channel Attention Weight Generation}
This richer multi-spectral descriptor $ \mathbf{Y}^{(i)}_{\text{dct}} $ is then fed into an attention weight generation block. This block typically consists of two fully connected (FC) layers forming a bottleneck structure (with a reduction factor $R$ and a \textit{rectified linear unit} (ReLU) activation after the first FC layer), followed by a final \textit{sigmoid} activation $ \sigma(\cdot) $. This process yields the channel attention weights:
\begin{equation} \label{eq:fcanet_attention_weights_redef}
     \mathbf{W}^{(i)}_{\text{att}} = \sigma( \text{FC}_2( \text{ReLU}( \text{FC}_1(\mathbf{Y}^{(i)}_{\text{dct}}) ) ) ) \in \mathbb{R}^{B \times L_c}.
\end{equation}

\subsection{Feature Rescaling}
Finally, the original input tensor $ \mathbf{H}_{c}^{(i)} $ (before any optional pooling) is recalibrated by element-wise multiplication ($ \odot $) with the generated attention weights $\mathbf{W}^{(i)}_{\text{att}}$, which can be expressed as:
% The weights are broadcast across the spatial dimensions $(N, E_c)$:
\begin{equation} \label{eq:feature_rescaling}
     \mathbf{H}_{c}^{(i+1)} = \mathbf{H}_{c}^{(i)} \odot \mathbf{W}^{(i)}_{\text{att}} .
\end{equation}
The result $\mathbf{H}_{c}^{(i+1)} \in \mathbb{R}^{B \times L_c \times N \times E_c}$ is the output of this attention layer, ready for the next layer or the output projection step.
\end{appendices}

\bibliographystyle{IEEEtran}
\bibliography{IEEEabrv,CRN}

\end{document}